\documentclass[preprint,12pt]{article}
\usepackage{authblk}

\usepackage{graphicx}
\usepackage{amsmath}
\usepackage{amsfonts}
\usepackage{amssymb}
\usepackage{mathptmx}      % use Times fonts if available on your TeX system
\usepackage{flushend}
\usepackage[colorlinks=true,linkcolor=red,citecolor=blue,urlcolor=blue]{hyperref}
\usepackage{color}
\usepackage{float}
\usepackage{xcolor}
\usepackage{enumitem}
\usepackage{subcaption}
\usepackage{multirow}
\usepackage{physics}
\usepackage[square, sort&compress, numbers]{natbib}
\usepackage{doi}
\begin{document}
\title{\bf{Universal relations for compact stars with exotic degrees of freedom}} 

\author[1]{Anil Kumar\thanks{\texttt{anil.1@iitj.ac.in}}}
\author[1]{Manoj Kumar Ghosh\thanks{\texttt{ghosh.9@iitj.ac.in}}}
\author[1]{Pratik Thakur\thanks{\texttt{thakur.16@iitj.ac.in}}}
\author[2]{Vivek Baruah Thapa\thanks{\texttt{vivek.thapa@bacollege.ac.in}}}
\author[3]{\thanks{Kamal Krishna Nath\texttt{kknath@niser.ac.in}}}
\author[1]{Monika Sinha\thanks{corresponding author: \texttt{ms@iitj.ac.in}}}
\affil[1]{\small Indian Institute of Technology Jodhpur, Jodhpur 342037 India}
\affil[2]{\small Department of Physics, Bhawanipur Anchalik College, Barpeta, Assam 781352, India}
\affil[3]{\small School of Physical Sciences, National Institute of Science Education and Research, An OCC of Homi Bhabha National Institute, Jatni 752050, India}

\maketitle
\begin{abstract}
The nature of the highly dense matter inside the supernova remnant compact star is not constrained by terrestrial experiments and hence modeled phenomenologically to accommodate the astrophysical observations from compact stars. The observable properties of the compact stars are highly sensitive to the microscopic model of highly dense matter. However, some universal relations exist between some macroscopic properties of compact stars independent of the matter model. We study the universal relation including the stars containing exotic degrees of freedom such as heavier strange and non-strange baryons, strange quark matter in normal and superconducting phases, etc. We examine the universal relations for quantities moment of inertia - tidal love number - quadrupole moment. We also study the correlation of non-radial f-mode and p-mode frequencies with stellar properties. We find the f-mode frequency observes the universal relation with dimensionless tidal deformability but the p-mode frequency does not show a good correlation with stellar properties. The p-mode frequency is sensitive to the composition of the matter. We find that universal relation is also applicable for stars with exotic matter in the core of the star with several models of exotic matter.
\end{abstract}
\section{Introduction}\label{sec:intro}
The massive stars end their lives by core-collapse supernova explosion leaving highly compact stars (CSs) as central objects. The CSs have an average density $\sim 10^{14}$ gm/cm$^3$ which is several times the normal nuclear saturation density ($n_0$). These compact objects accommodate the extremely dense matter in the universe and thus allow insights into various new aspects of dense matter physics. Such high densities are not attainable in any laboratory or terrestrial experiments and hence the exact composition of matter as well as the inter-particle interaction inside the compact objects is unknown. The equation of state (EOS), which describes the relationship between energy density ($\varepsilon$) and pressure (p), at extremely high densities, is among the most uncertain aspects of nuclear physics. Consequently, many phenomenological models are discussed with different compositions and different inter-particle interactions. The most studied and discussed possibility of matter at such high density is pure nucleonic matter - the matter composed of mostly neutrons with some admixture of protons and electrons. However, the appearance of exotic degrees of freedom in the interior of massive NSs where the density is a few times $n_0$ tends to be feasible, although it remains an open question. For example, there are possibilities of the appearance of strange baryons \cite{1991PhRvL..67.2414G,2013PhRvC..87e5806C, 2015JPhG...42g5202O, 2018MNRAS.475.4347R,2018EPJA...54..133L,2021NuPhA100922171L}, non-strange heavier baryons \cite{2014PhRvC..90f5809D, 2015PhRvC..92a5802C, 2018PhLB..783..234L, 2019ApJ...874L..22L, 2023ApJ...944..206L, thapa2020equation, 2021MNRAS.507.2991T}, Boson condensates \cite{2019Parti...2..411M, 1982ApJ...258..306H, 1999PhRvC..60b5803G, 2001PhRvC..63c5802B, 2020PhRvD.102l3007T, thapa2021massive} etc. Another possibility is that at much higher density, the baryons, may get decomposed into their constituent quarks and generate a region of deconfined strange quark matter (SQM) \cite{collins1975superdense,NKGb_1997}. The star may be entirely composed of SQM up to the surface due to some stability conditions \cite{1971SvA....15..347S, PhysRevD.30.272,2022MNRAS.513.3788K}. Another possibility is a hybrid star (HS) composed of SQM at the core surrounded by baryonic matter up to the surface \cite{2018ApJ...857...12N, 2019ApJ...877..139G, 2019MNRAS.489.4261M, 2021PhRvC.103e5814R, 2023PhRvD.107f3024K}. 

The CSs have an interior structure that is highly sensitive to their EOS. Consequently, these factors dictate the external characteristics of the objects in question, including their mass and radius; their deformability, as indicated by their quadrupole moment $Q$ and tidal Love number $\Lambda$; and their rotation rate, which is defined by their moment of inertia $I$. The macroscopic properties of CSs are highly sensitive to the microscopic properties of highly dense matter which are still unknown and model-dependent as of now. Hence, the only way to minimize the theoretical uncertainties regarding highly dense matter is to fit the proposed models with the astrophysical observations coming from these CSs.  

However, Yagi and Yunes in 2013 discovered that some combinations of physical parameters do not depend on the EOS and follow universal relations \cite{2013Sci...341..365Y,2013PhRvD..88b3009Y}. They showed that the relations between the moment of inertia (I), tidal Love number ($\Lambda$), and quadrupole moment (Q) of CSs are independent of EOS of highly dense matter irrespective of whether the matter is pure nucleonic or deconfined strange quark matter. The combinations of the above-mentioned variables obey a universal relation. Recently, universal relations have been getting some attention \cite{2016CQGra..33mLT01Y, 2018PhRvD..97h4038P,2018CQGra..35a5005S,2018PhRvD..97f4042G,2017PhRvC..96d5806M,2019JPhG...46c4001W,2013PhRvD..88b3007M,2013ApJ...777...68B,2014MNRAS.438L..71H,2014PhRvL.112l1101P,2014PhRvD..89l4013Y,2014PhRvL.112t1102C,2021MNRAS.502.3476S,2019PhRvD..99d3004R,2020PhRvD.101l4006J,2020MNRAS.499..914R,2020PhRvC.101a5805K,2021PhRvD.103f3036G,2021ApJ...906...98N,2022MNRAS.515.3539K,2022PhRvD.106l3002Z,2023MNRAS.524.1438N,2018EPJA...54...26B}. Moreover, later on some empirical relations of non-radial oscillation frequencies with stellar parameters and universal relation have been reported \cite{1998MNRAS.299.1059A, 2004PhRvD..70l4015B, 2015PhRvD..91d4034C,2023PhRvD.107b3010P}.

Universal relations are crucial because they enable us to calculate the others if we know one parameter. When it comes to CSs, it is possible to compute the love number $\Lambda$ and $Q$ without physically measuring them if $I$ can be measured in some way. This is extremely useful, as both are challenging to quantify for binaries separated by a significant distance. By accurately measuring the love number in coalescing binaries, it is possible to infer the $Q$ and, consequently, gain insight into the spins of the coalescing binary CSs. Most of the studies regarding universal relations have been carried out with neutron stars composed of pure nucleonic matter. The $I-Q$ relation for rapidly rotating NSs was first studied by Doneva \cite{2014ApJ...781L...6D}. They found that the $I-Q$ relation is broken and becomes more EOS-dependent for NSs with a fixed frequency. However, it was soon found by Pappas \& Apostolatos \cite{Pappas_2014}, and Chakrabarti et al. \cite{2014PhRvL.112t1102C} that the $I-Q$ relation can remain approximately EOS-insensitive if one chooses suitable dimensionless parameters instead of dimensional quantities. Bandyopadhyay et al. \cite{2018EPJA...54...26B} studied the universal relations with $\Lambda$-hyperons and anti-kaon condensates. Eemeli Annala et al. \cite{2018JHEP...12..078A} reported a maximum deviation of $15\%$ to $20\%$ in universal relations due to strong first-order phase transition at low density. Raduta et al. \cite{2020MNRAS.499..914R} discussed the universal relation considering hot stars with heavier strange and non-strange baryons and quarks in the core.  The universal relation for hot and cold rapidly rotating HSs has been considered by Largani et al. \cite{2022MNRAS.515.3539K}. The implication of GW170817 with universal relations for HS has been studied by Paschalidis et al. \cite{2018PhRvD..97h4038P}. %considering Maxwell construction  
At high enough density ($\mu>>T$), there may exist the Color-Flavor-Locked (CFL phase), which is a superconducting phase of SQM \cite{2008RvMP...80.1455A, 2005PrPNP..54..193W, 2002PhRvD..66i4007S}.   
%Recently, the universality of HS in various temporal and rotational conditions has been studied in refs \cite{2018PhRvD..97h4038P, 2022MNRAS.515.3539K}. 
In recent works, we extend the study of universalities in the case of CS with strange and heavier non-strange baryons as well as SQM in the Color-Flavor-Locked (CFL) phase inside the interior of the star along with the pure NS. 

There are several quasi-normal modes like the fundamental f-mode, pressure p-modes, gravity g-modes, spacetime w-modes, etc. \cite{1999LRR.....2....2K, lindblom1983quadrupole, detweiler1985nonradial}, each classified based on the restoring forces that act to bring it back to equilibrium. For example, the f- and p-modes, which are acoustic waves in the star, are restored by fluid pressure while g-modes which arise due to discontinuities in density or temperature are restored by gravity (buoyancy). Various previous works have explored non-radial modes in cold and finite temperature neutron stars \cite{2022PhRvD.106f3005K, zhaoUniversalRelationsNeutron2022, lozanoTemperatureEffectsCore2022, thapaFrequenciesOscillationModes2023}. Several years ago, Andersson and Kokkotas \cite{1998MNRAS.299.1059A}  put forward an empirical relation for the f-mode oscillation frequency $\omega$ with the stellar parameters mass and radius, based on the Newtonian theory of stellar perturbations. They observed that in full GR, $\omega$ depends almost linearly on the square root of the average density. Another relation, based on estimates using the quadrupole formula, was established for the damping time due to gravitational wave emission, $\tau$. Later, Benhar et al. \cite{2004PhRvD..70l4015B} presented further results that included more and newer equations of state, updating the fits from \cite{1998MNRAS.299.1059A}. The average frequencies $\omega$ were systematically lower than the one for the old EOS sample, which they attributed to the fact that the new sample included stiffer EOSs. Later \cite{2015PhRvD..91d4034C} thoroughly studied the universality. They considered a wide range of masses and the EOSs to minimize the uncertainty. In this work, we took it further and computed the universality for non-radial oscillation frequencies for all the possible families of CSs. In this work, we adopt the Cowling classification, where the various modes are separated by the number of radial nodes \cite{coxNonradialOscillationsStars1976,rodriguezThreeApproachesClassification2023}. The f-mode, whose frequency lies between those of the p- and g-modes, has no radial node number.
Quadrupolar oscillations (l=2) of all modes lead to the emission of GWs. With the advent of enhanced next-generation telescopes like the Cosmic Explorer and the Einstein telescope which carry about 10 times the sensitivity of Advanced LIGO, the possibility of detection of these modes increase \cite{2001MNRAS.320..307K}.  Although the high-frequency p-modes aren't expected to be detected by the next-generation GW detectors, we include their study for the sake of completeness. 

In the Sect. \ref{sec:star}, we describe the models of highly dense matter considered in the work and structure of stars composed of matter within these models. In Sect. \ref{sec:non-radial}, we discuss the non-radial modes and determine their frequencies for CS with heavier baryons and SQM in the CFL phase inside the core. Next, in Sect. \ref{sec:universal} we study the universal relations between different quantities.

\section{Matter model and consequent stellar structure}\label{sec:star}
In this work, we study the universal relation for different observable quantities related to CS composed of matter with different possible compositions. As mentioned in the Sect. \ref{sec:intro}, the most discussed stellar model is with nucleonic matter. With this, we consider the appearance of heavier strange and non-strange baryons and SQM in the core of the star. In this context, first, we briefly discuss the star with baryonic matter only in the subsequent subsection. Then, in the next subsection, we discuss the HS with SQM in normal and CFL phases inside the core.

\subsection{Baryonic star}\label{subsec:bs}
Inside CS highly dense matter is thought to be composed of mostly neutrons with a small admixture of protons and electrons. With the advent of astrophysical observation, now it is established that the CS can be as massive as $2~M_\odot$ or even more than that which leads to the assumption of the existence of matter at several times nuclear saturation density near the core of the CS. In that case, near the core, there is a possibility of the appearance of heavier baryons. We consider the Relativistic Mean Field (RMF) approach for baryonic matter at high density. Within the RMF model, we assume the interaction between nucleons is mediated by the exchange of isoscalar-scalar $\sigma$, isoscalar-vector $\omega$ and $\phi$, and isovector-vector $\rho$ mesons. The recent observations of massive CS candidates indicate that the density inside the core is much higher than $n_0$ with which the appearance of heavier strange and non-strange baryons is highly probable. On the other hand, the appearance of heavier baryons softens the matter reducing the maximum attainable mass of CS which contradicts the recent astrophysical observations. To come out of this problem, one should consider the baryon interaction density-dependent. Hence, in our current work, we stick to that formalism within the RMF approach. We consider the matter composed of only nucleons as well as the possibility of the appearance of strange baryons belonging to baryon octet and $\Delta$-resonances \cite{2021MNRAS.507.2991T}.

In the RMF approach, the strength of baryon coupling to mesons is taken as a parameter. The values of the parameters are fixed from observed nuclear matter properties at $n_0$, such as values of $n_0$, binding energy per nucleon $E/A$ at $n_0$, symmetry energy $E_{sym}$ at $n_0$ and its derivative with respect to $n_0$, incompressibility $K$, etc. Different sets of parameter values can reproduce these observed nuclear matter properties. Depending on this several models of parametrizations are there for baryonic matter models. In this work, we employ the parameterizations namely DD2 \cite{PhysRevC.81.015803}, DD-ME2 \cite{2005PhRvC..71b4312L} and DD-MEX \cite{TANINAH2020135065} for baryonic matter which covers the nuclear saturation properties as shown in Table \ref{tab:sat}. 
\begin{table*}
\centering
\resizebox{\textwidth}{!}{\begin{tabular}{|l|l|l|l|l|l|}
\hline
\begin{tabular}[c]{@{}l@{}}n$_0$\\ $(1/fm^3)$\end{tabular} & \begin{tabular}[c]{@{}l@{}}-E$_0$\\ (MeV)\end{tabular} & \begin{tabular}[c]{@{}l@{}}K$_0$\\ (MeV)\end{tabular} & \begin{tabular}[c]{@{}l@{}}E$_{sym}$\\ (MeV)\end{tabular} & \begin{tabular}[c]{@{}l@{}}L$_{sym}$\\ (MeV)\end{tabular} & \begin{tabular}[c]{@{}l@{}}-K$_{sym}$\\ (MeV)\end{tabular} \\ \hline
0.149 - 0.152                                       & 16.02 - 16.14                                      & 242.70 - 267.06                                   & 32.269 - 32.30                                       & 49.576 - 54.96                                       & 71.47 - 93.24                                         \\ \hline
0.17 $\pm$ 0.03 \cite{2012PhRvC..85c5201D}                                       & $\sim 16$ \cite{2014PhRvC..90e5203D}                                      & $210-280$ \cite{2017RvMP...89a5007O}                                   & $28.5-34.9$ \cite{2017RvMP...89a5007O}                                       & $30.6-86.8$ \cite{2017RvMP...89a5007O}                                       & $39.7-182.3$ \cite{2017RvMP...89a5007O}                                         \\ \hline
\end{tabular}}
\caption{The range of nuclear parameters at n$_0$. The lower row represents the values (along with references) of the saturation properties obtained from various terrestrial experiments and methods.}
\label{tab:sat}
\end{table*} 
The readers may refer to Ref. \cite{2017RvMP...89a5007O} for further details of the comparison of experimental data from finite nuclei and heavy-ion collisions with different microscopic model calculations.

The equation of state (EOS) for only nucleon matter is shown by the curves with circles in Fig. \ref{fig:EOS_MR}. When we consider the possibility of the appearance of strange baryons (hyperons) then the EOS softens as shown in the curves with the thin dotted line. The appearance of non-strange heavier baryons $\Delta$ further softens the EOS at low density and affects the maximum attainable mass as evident from curves with the thick dotted line in the same figure.

\subsection{Hybrid star}
With the increase in density towards the center, the deconfinement of quarks is another possibility that leads to the formation of SQM at the core of the star, making an HS. Naturally, inside HS, the deconfined SQM at the core is surrounded by pure hadronic matter. In our recent work we consider the outer baryonic part of the HS to be purely nucleonic. Inside the core for quark matter, we consider two possibilities: one is normal SQM matter and another possibility is superconducting quark matter in the CFL phase. Various models are available for the description of SQM. However, the MIT bag model and NJL model are well established. The original MIT bag model considers free quarks confined within a bag. Subsequently, some interaction between the quarks has been incorporated \cite{NKGb_1997,1984PhRvD..30.2379F}. In our work we consider the MIT bag model with interaction and denote it as MIT bag model. In another model, repulsive vector interaction terms are included within free quark MIT bag model which is known as vector bag model (vBAG) \cite{2022MNRAS.513.3788K,lopes2021modified,2016MNRAS.463..571F,2015ApJ...810..134K}. For the superconducting CFL phase, we use the MIT bag model including the gap term as mentioned in Ref. \cite{2003A&A...403..173L,2022JPhG...49g5201S,2020IJMPD..2950044R}. For the universal relations we consider the inner core SQM inside the HS considering these models mentioned above, which are briefly discussed in the subsequent sections.

\subsubsection{Normal quark matter}
The normal SQM inside the core of the HS is composed of up(u), down(d), strange(s) quarks, and electrons(e).

\begin{figure}[t]
\centering
\includegraphics[width=0.7\textwidth]{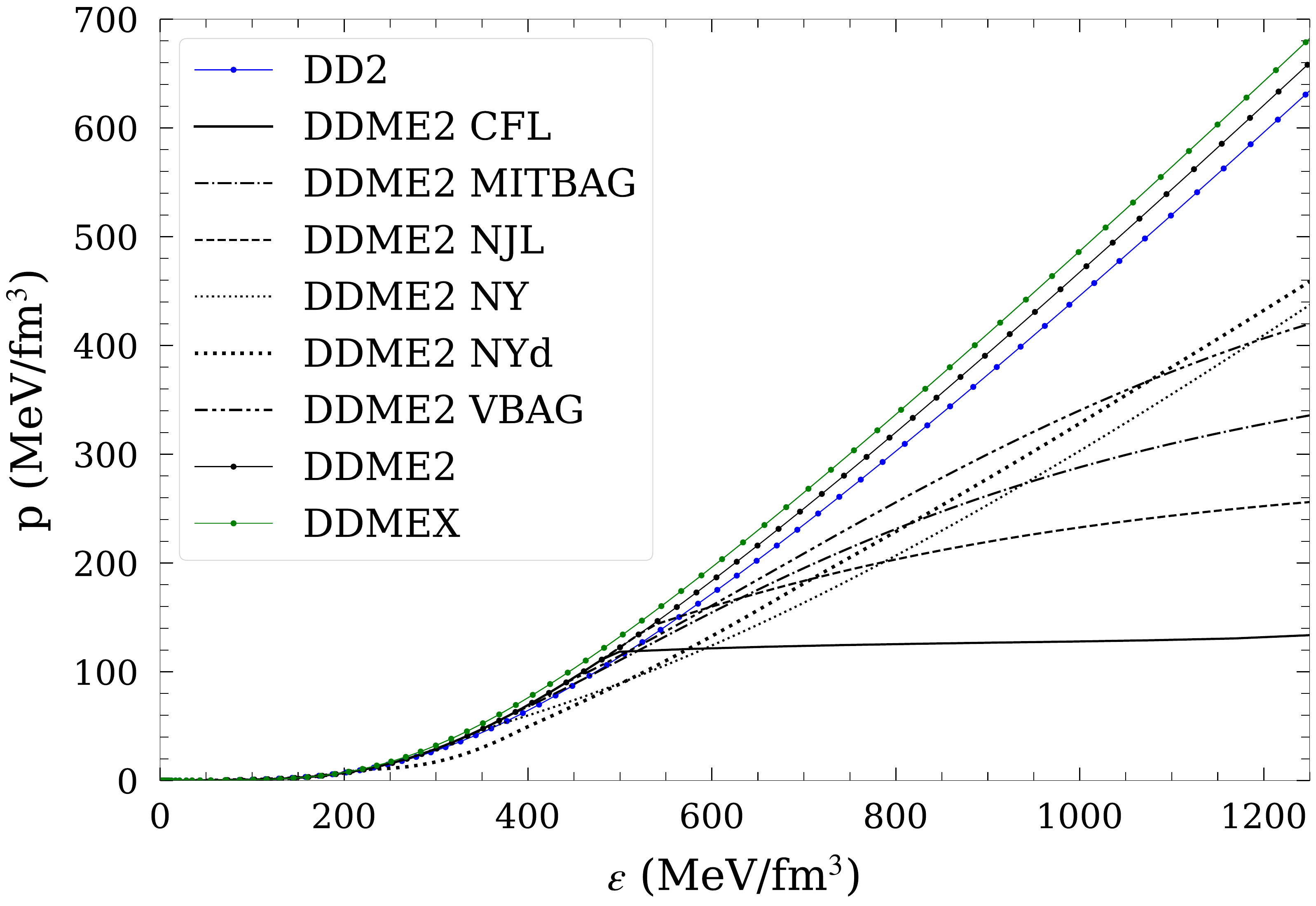}
\caption{The variation of pressure with energy density}
\label{fig:EOS_MR}
\end{figure}

\paragraph{ MIT bag model:} 
Here we briefly describe the MIT bag model with interaction as introduced by Farhi and Jaffe \cite{1984PhRvD..30.2379F}. In this model
the thermodynamic potential of each quark is given by
\begin{equation}
\begin{split}
   \Omega_q & = -\frac{1}{4\pi^2}\Biggl\{\mu_q\sqrt{\mu_q^2 - m_q^2}\left(\mu_q^2-\frac{5}{2}m_q^2\right)\\
   & +\frac{3}{2}m_q^4\ln{\frac{\mu_q+\sqrt{\mu_q^2-m_q^2}}{m_q}} \\ 
   & - \frac{2\alpha_s}{\pi}\Biggl[3\left(\mu_q\sqrt{\mu_q^2-m_q^2}-m_q^2\ln{\frac{\mu_q+\sqrt{\mu_q^2-m_q^2}}{m_q}}\right)^2 \\
   & -2\left(\mu_q^2 - m_q^2\right)^2 -3m_q^4\ln^2{\frac{m_q}{\mu_q}} \\
   & + 6\ln{\frac{\sigma}{\mu_q}}\left(\mu_qm_q^2\sqrt{\mu_q^2 - m_q^2} - m_q^4\ln{\frac{\mu_q+\sqrt{\mu_q^2-m_q^2}}{m_q}}\right) \Biggr]\Biggr\}
\end{split}
\end{equation}
with $\mu_q$ and $m_q$ the chemical potential and mass of each quark $q$ respectively, $\alpha_s$ the coupling parameter for strong interaction and $\sigma$ the renormalization scale. 
The thermodynamic potential of the electron is 
\begin{equation}
    \Omega_e = - \frac{\mu_e^4}{12\pi^2}
\end{equation}
with $\mu_e$ the electron chemical potential.
Then the EOS can be obtained through the relations
\begin{equation}
\varepsilon = \sum_{q=u,d,s}(\mu_qn_q + \Omega_q) + \Omega_e + B
\end{equation}
\begin{equation}
p = \sum_{i=u,d,s,e}\mu_in_i - \varepsilon
\end{equation}
Here, $B$ is the bag pressure that separates the quarks confined within the bag from the vacuum.
where
\begin{equation}
    \mu_i = \sqrt{k_i^2+m_i^2}
\end{equation}
and the number density is
\begin{equation}
    n_i = -\frac{\partial\Omega_i}{\partial\mu_i}
\end{equation}
We consider two sets of parameters with this model one is $\alpha_s = 0.8$ fm$^2$ with $B^{1/4} = 160$ MeV and another is for softer EOS $\alpha_s = 0.3$ fm$^2$ with $B^{1/4} = 174$ MeV. With this quark model, we consider the up quark mass $m_u = 4$ MeV, down quark mass $m_d = 7$ MeV, and strange quark mass $m_s = 100$ MeV. The value of renormalization is $300$ MeV here. The appearance of SQM naturally softens the EOS. With a small strength of interaction parameters and a high value of bag parameter, the EOS is comparatively softer. The stiffer EOS of HS is shown by the dash-dotted line in Fig. \ref{fig:EOS_MR} with this model.

\paragraph { vBAG model:}
The Lagrangian density of SQM with the vBAG model is given by the following equation as
\begin{equation}
\begin{aligned}
    {\cal L}_Q & = \sum_{q = u,d,s}\left[\bar{\psi}_q\{\gamma_{\mu}\left(i\partial^\mu - g_{qV}V_\mu\right)-m_q\}\psi_q-B \right] \Theta(\bar{\psi}_q\psi_q) \\ & - \frac{1}{4}{\left(\partial_{\mu}V_{\nu}-\partial_{\nu}V_{\mu}\right)}^2 + \frac{1}{2}{m^2_V}{V_\mu}V^{\mu} +\bar{\psi}_e\left(\gamma_{\mu}i\partial^\mu -m_e\right)\psi_e 
\end{aligned} 
\end{equation}
where $\psi_q$ is the field of quark q considering it as a fermion and $m_q$ is its mass. Similarly, $V_\mu$ and $m_V$ are the field of the mediator and its mass respectively. $\Theta$ is the heavy side function that is unity inside and zero outside the bag. After solving equations of motions for the Lagrangian density, the chemical potential gets shifted as
\begin{equation}
    \mu_q = \sqrt{(k_{fq})^2 + (m_q)^2} + g_{qV}V_0
\end{equation}
The energy density of SQM matter within the vBAG model is 
\begin{equation}
\begin{aligned}
\varepsilon & = \frac{3}{\pi^2}\sum_{q}\int_{0}^{k_{f_q}}\left(\sqrt{(k)^2 + (m_q)^2}+ g_{qV}V_0\right)k^2dk \\ & - \frac{1}{2}\left({m_V}{V_0}\right)^2 + B,
\end{aligned}
\end{equation}
here $V_0$ is the mean value of $V_\mu$ in ground state. From energy density, pressure can be obtained by the following equation
\begin{equation}
    p = \sum_{q}\mu_qn_q - \varepsilon
\end{equation}
where $n_q$ signifies number density of quark $q$. For simplicity, we define the parameter $G_V = (g_{qv}/m_V)^2$. In this work, we take two sets of parameters, one is a higher value of this parameter $G_V = 0.25$ fm$^2$ with $B^{1/4} = 180$ MeV and another is a lower value $G_V = 0.17$ fm$^2$ with $B^{1/4} = 172$ MeV for softer EOS. With the vBAG model, we consider masses quarks the same as the MIT bag model. The inclusion of vector interaction stiffens the EOS compared to the MIT bag model. For HS with SQM core considering this model, the stiffer EOS is shown by the dash double dotted line in Fig. \ref{fig:EOS_MR} with DDME2 nuclear EOS.

\paragraph{ NJL model: }
The Lagrangian density describing the dynamics of SQM within the NJL model with vector interactions and the 't Hooft six-quark interaction term is given by \cite{2022Ap.....65..278A,2005PhR...407..205B,PhysRevLett.37.8,2022PhRvC.105c5802P}:
\begin{equation}
\begin{aligned}
\mathcal{L} &= \bar{\psi}(i\gamma^{\mu}\partial_{\mu} - \hat{m_0})\psi + G_S\sum_{a=0}^{a=8}\left[(\bar{\psi}\lambda_a\psi)^2 + (\bar{\psi}i\gamma_5\lambda_a\psi)^2\right] \\
&\quad - G_V\sum_{a=0}^{a=8}\left[(\bar{\psi}\gamma_\mu\lambda_a\psi)^2 + (\bar{\psi}i\gamma_\mu\gamma_5\lambda_a\psi)^2\right]\\
&\quad + K\left[\text{det}_f(\bar{\psi}(1-\gamma_5)\psi) + \text{det}_f(\bar{\psi}(1+\gamma_5)\psi)\right] \\
&\quad + \bar{\psi_e}(i\gamma^{\mu}\partial_{\mu} - m_e)\psi_e
\end{aligned}
\end{equation}
where $\lambda_a$ are Gell-Mann matrices. $\psi=(u,d,s)^T$ is fermionic field of quarks and $\hat{m_0}$ is diagonal matrix with elements $(m_{u0},m_{d0},m_{s0})$. The scalar interaction and 't Hooft interaction strengths are denoted by the parameters $G_S$ and $K$ respectively. In Fig. \ref{fig:EOS_MR}, the HS EOS with NJL model for quark matter is represented by the dashed line. 
\begin{equation}
\begin{aligned}
\varepsilon &= \frac{3}{\pi^2} \sum_{f=u,d,s} \left[\int_{0}^{\Lambda}dkk^2\sqrt{k^2+M_{f0}^2} - \int_{k_{F_f}}^{\Lambda}dkk^2\sqrt{k^2+M_{f}^2}\right] \\
&\quad + 2G_S(\sigma_u^2 + \sigma_d^2 + \sigma_s^2 - \sigma_{u0}^2 - \sigma_{d0}^2 - \sigma_{s0}^2)\\
&\quad - 4K(\sigma_u \sigma_d \sigma_s - \sigma_{u0} \sigma_{d0} \sigma_{s0})+2G_V(n_u^2 + n_d^2 + n_s^2)\\
&\quad + \frac{1}{\pi^2}\int_{0}^{K_{F_e}}dkk^2\sqrt{k^2+m_e^2},
\end{aligned}
\end{equation}
Here, the constituent quark mass $M_f$ and the constituent quark mass for vacuum ($n_u = n_d = n_s = 0$) $M_{f0}$ are given by 
\begin{equation}
    M_f = m_{f0} - 4G_S\sigma_f + 2K\sigma_i\sigma_j
\end{equation}
and
\begin{equation}
    M_{f0} = m_{f0} - 4G_S\sigma_{f0} + 2K\sigma_{i0}\sigma_{j0}
\end{equation}
respectively with the chiral condensates
\begin{equation}
\begin{aligned}
    \sigma_f = - \frac{3M_f}{\pi^2}\int_{k_{F_f}}^{\Lambda}dk\frac{k^2}{\sqrt{k^2+M_{f}^2}} \\
    \sigma_{f0} = - \frac{3M_{f0}}{\pi^2}\int_{0}^{\Lambda}dk\frac{k^2}{\sqrt{k^2+M_{f0}^2}}    
\end{aligned}    
\end{equation}
We consider the mass of quarks as $m_{u0}=m_{d0}=5.5$ and $m_{s0}=140.7$ MeV. Value of ultraviolet cutoff parameter in momentum space ($\Lambda$)
is considered $602.3$ MeV. Parameters $G_S$ and $K$ are related to $\Lambda$ as $G_S=1.835/\Lambda^2$ and $K=12.36/\Lambda^5$ respectively. The vector interaction parameter is $G_V=0.2G_S$.

\subsubsection{Superconducting quark matter}
In this phase, quarks are paired up in such a way that they form color-neutral and flavor-neutral condensates, and this leads to a unique and stable ground state of matter. In such a scenario, these paired quarks would move together with correlated momentum, somewhat analogous to Cooper pairs in superconductors. Studying the CFL phase in neutron stars is intimately linked to quantum chromodynamics (QCD), the theory of strong interactions. 
Here, each quark has a common Fermi momentum ($\nu$) and number density ($n$) to form Cooper pairs
\begin{equation}
\nu = 2\mu - \sqrt{\mu^2+\frac{{m_s}^2}{3}}
\end{equation}
\begin{equation}
n = \frac{\nu^3+2\Delta^2\mu}{\pi^2}   
\end{equation}
Here, $\Delta$ is the CFL energy gap and quark number chemical potential is $\mu = (\mu_u+\mu_d+\mu_s)/3$. The thermodynamic potential of this phase is given by
\begin{equation}
    \Omega_{CFL} = \Omega_{free} - \frac{3}{\pi^2}\Delta^2\mu^2 + B
\end{equation}
with
\begin{equation}
    \Omega_{free} = \frac{6}{\pi^2}\int_{0}^{\nu}(p-\mu)p^2dp + \frac{3}{\pi^2}\int_{0}^{\nu}(\sqrt{p^2+{m_s}^2}-\mu)p^2dp    
\end{equation}
The pressure ($p$), and energy density ($\varepsilon$) of CFL phase are
\begin{equation}
    p = -\Omega_{CFL}
\end{equation}
\begin{equation}
    \varepsilon = \sum_{i=u,d,s}\mu_in_i + \Omega_{CFL} 
\end{equation}
The quark matter EOS with the CFL phase is represented by the solid line.
With these models of SQM we construct the core of the HS surrounded by nucleonic matter with the models discussed in the Subsect. \ref{subsec:bs}. To construct the HS we consider Gibb's construction in which equilibrium between the baryonic phase and quark phase is established in a mixed phase during the phase transition. 
In this phase, global charge neutrality is maintained rather than local charge neutrality as
\begin{equation}
    \chi\rho_c^{Q} + (1-\chi)\rho_c^{B} = 0,
\end{equation}
where $\chi$ is the volume fraction of quark matter to baryonic matter and $\rho_c$ is the charge density. Superscripts B and Q denote the baryonic phase and quark phase respectively. The energy density of mixed-phase is given by the following equation
\begin{equation}
    \varepsilon_{MP} = \chi\varepsilon_{Q} + (1-\chi)\varepsilon_{B}
\end{equation}
here $\varepsilon_Q$ and $\varepsilon_H$ are the energy densities of quark matter and baryonic matter respectively. The baryonic density of matter in the mixed phase is 
\begin{equation}
    \rho_{MP} = \chi\rho_{Q} + (1-\chi)\rho_{B}
\end{equation}
Baryonic matter density for quark matter is $\rho_Q = (n_u+n_d+n_s)/3$. The pressure of mixed-phase is equal to the pressure of quark matter and baryonic matter.
With this, we construct the HS structure with different combinations of the baryonic matter model and SQM models as discussed above.

\begin{table}[h]
\begin{center}
\begin{tabular}{cccccccccccc}
\hline \hline
      B.M. & Q.M. &$B^{1/4}$& & \\
      & & (MeV) & & \\         
\hline
  DD2 & CFL & 190 & $\Delta=30$ (MeV) \\
  DDMEX & CFL & 205 & $\Delta=15$ (MeV) \\
  DDME2 & CFL & 205 & $\Delta=30$ (MeV) \\
  DD & vBAG & 180 & $G_V=0.25$ ($fm^2$)\\
  DD & vBAG(s) & 172 & $G_V=0.17$ ($fm^2$)\\
  DD & MITbag & 160 & $\alpha_s=0.8$\\
  DD & MITbag(s) & 174 &  $\alpha_s=0.3$\\
\hline
\end{tabular}
\end{center}
\caption{Set of parameters we used for HSs EOS, B.M. denotes baryonic matter and Q.M. denotes quark matter}
\label{tab:eos}
\end{table}

\subsection{Comparison of the models}
It is well known that nuclear matter is the stiffest among all possible compositions of matter. The appearance of exotic matter at higher density softens the matter undoubtedly. From the comparison of the appearance of different exotic degrees of freedom, we notice that the appearance of the quark softens more compared to the hyperon appearance. That is evident from the Fig. \ref{fig:EOS_MR}. Here we show the nuclear matter with three parametrizations as discussed in sec. \ref{subsec:bs}. We see that DDMEX is the stiffest and DD2 is the softest among them for nuclear matter. To understand the effect of exotic matter, we consider the nucleonic part of the matter with DDME2 parametrization.
When considering the appearance of hyperonic matter it makes the EOS softer. In the case of hyperonic matter with $\Delta$ baryons, $\Delta$ starts appearing very early and makes matter soft at a low-density regime. Without $\Delta$ other strange baryons appear a little later. It also softens the matter more at high density. SQM appears at comparatively higher density depending on the model and model parametrizations considered. The stiffness of the matter decreases progressively from the vBAG model to the MITBAG model, then further to the NJL model, culminating in the CFL phase. In very high-density regions matter with SQM is softer than the matter with hyperons. The parametrization for this plot is detailed in Table \ref{tab:eos}. These models incorporate a broad range of soft to stiff EOSs, making them ideal for testing the robustness of our universal relations.
In Fig. \ref{fig:EOS_MR}, every EOS has central energy density corresponding to maximum mass less than $1250$ $\text{MeV}/{\text{fm}^3}$ and pure SQM does not appear inside the core of HS. So, for HS we include softer EOSs with vBAG and MITBAG models. The different combinations of parameters are given in Table \ref{tab:eos}. In this table, we denote the softer parameter sets within a model with the notation $(s)$ beside the name of the models.

\section{Non-radial modes}\label{sec:non-radial}
Now we describe the methods of finding the frequencies of non-radial modes relevant for the CSs. The fully general relativistic calculation is somewhat cumbersome, the most used approximation is the Cowling approximation. We have used that as well as the full general relativistic calculation to find the non-radial frequencies.

\subsection{Relativistic Cowling approximation}
First, we describe the Cowling approximation. In this approximation, the metric is assumed to be unperturbed. Since the non-radial oscillation formalism is developed in linearised gravity, the line element of a non-rotating neutron star is taken as:
\begin{equation}
 ds^2 = -e^{2\Phi(r)}dt^2 + e^{2\Lambda(r)}dr^2 + r^2d\theta^2 + r^2\sin^2{\theta}d\phi^2   
\end{equation}

The fluid Lagrangian displacement vector is assumed to be: 
\cite{2011PhRvD..83b4014S}
\begin{equation}
\begin{aligned}
\xi^i = \left(e^{-\Lambda}W,-V\partial_\theta,-V\sin^{-2}\theta\partial_\phi\right)r^{-2}\mathcal{Y}_{lm}(\theta,\phi) 
\end{aligned}
\end{equation} 
where $\mathcal{Y}_{lm}(\theta,\phi)$ are the spherical harmonics. $W$ and $V$ are perturbative fluid variables, which are functions of $r$ with a harmonic time dependence $W(t,r)=W(r)e^{i{\omega}t}$ and $V(t,r)=V(r)e^{i{\omega}t}$. The mode frequencies $\omega$ are thus found by solving the following system of ordinary differential equations \cite{2011PhRvD..83b4014S}: 

\begin{equation}
\begin{aligned}
W' &= \dv{\varepsilon}{p}\left[{\omega}^2r^2e^{\Lambda-2\Phi}V + \Phi'W\right] - l(l+1)e^{\Lambda}V\\    
V' &= 2\Phi'V-r^{-2}e^{\Lambda}W
\end{aligned}\label{eqn-mode osc}    
\end{equation}
where $\varepsilon$ is the energy density and $p$ the pressure. The dash ($'$) represents a derivative with respect to the radius. The boundary conditions at the center of the star are $W(r\to0)= Ar^{l+1}$ and $V(r\to0)=-Ar^l/l$, where $A$ is an arbitrary constant, usually taken to be $1$. 

The surface boundary condition corresponds to the fluid pressure vanishing at the surface of the star ($r= R$): 
\begin{equation}
{\omega}^2e^{\Lambda(R)-2\Phi(R)}V(R) + \frac{1}{R^2}\frac{d\Phi(r)}{dr}\bigg\vert_{r=R}W(R) = 0 \label{eqn-surf cond}    
\end{equation}

The lowest frequency solution to the oscillation mode equation has no radial node and thus corresponds to the f-mode frequency. The next highest solution has one radial node and is thus the first p-mode (p1-mode). Although \cite{rodriguezThreeApproachesClassification2023} suggests that this classification fails before 0.4 seconds of the bounce, for the models in consideration in this work, the Cowling classification holds good. The number of radial nodes was found by counting the number of times the perturbative variables $W$ and $V$ become $0$ within $r<R$ since this condition ensures that the three-velocity of the fluid (which is the time derivative of the fluid Lagrangian displacement vector), becomes $0$ at those points. The correctness of our code was ensured when our results matched with those reported in \cite{2022PhRvD.106f3005K,thapaFrequenciesOscillationModes2023}.

\subsection{Fully general relativistic calculations}
In fully general relativistic calculations, we consider the perturbation in metric too.
The equations that need to be solved and the various techniques required to solve them have been examined in great detail in several previous works \cite{lindblom1983quadrupole,detweiler1985nonradial,zhaoUniversalRelationsNeutron2022,1969ApJ...158....1T,chandrasekhar1991,sotaniDensityDiscontinuityNeutron2001} and we simply state them for completeness. 

In this work, we follow the method of direct numerical integration, first proposed by \cite{lindblom1983quadrupole,detweiler1985nonradial} and refined by \cite{lujunli_ChinPhyB}. The perturbed metric is taken as:
\begin{align}
    ds^2&= -e^{2\Phi(r)}(1+r^lH_0(r)\mathcal{Y}_{lm}e^{i\omega t})dt^2- 2i\omega r^{l+1}H_1(r)\mathcal{Y}_{lm}\nonumber\\
    &\quad \cross{e^{i\omega t}dt dr}\nonumber\\
    &\quad + e^{2\Lambda(r)}(1-r^lH_0(r)\mathcal{Y}_{lm}e^{i\omega t})dr^2\nonumber\\
    &\quad + r^2(1-r^lK(r)\mathcal{Y}_{lm}e^{i\omega t})(d\theta^2+ \sin^2\theta \ d\phi^2) 
\end{align}
\newpage
while the fluid displacement vector in terms of the perturbation functions $W(r)$ and $V(r)$ is: 
\begin{equation}
\begin{aligned}
\xi^i &= \left(r^{l-1}e^{-\Lambda}W(r),-r^{l-2}V(r)\partial_\theta,-r^{l-2}\sin^{-2}\theta V(r)\partial_\phi\right)\mathcal{Y}_{lm}(\theta,\phi)\\
&\times{e^{i\omega t}}
\end{aligned}
\end{equation}
Lindblom and Detweiller \cite{lindblom1983quadrupole,detweiler1985nonradial} introduced a new function $X(r)$ to replace $V(r)$ and the relations between all these functions are as follows: 
\begin{subequations}
    \begin{align}
        H_0 &= \bigg\{8\pi r^3e^{-\Phi}X- \bigg[\frac{1}{2}l(l+1)(m+4\pi r^3 p) - \omega^2 r^3 e^{-2(\Lambda+\Phi)}\bigg]H_1\nonumber\\
        &\quad + \left[\frac{1}{2}(l+2)(l-1) r- \omega^2r^3e^{-2\Phi} - \frac{e^{2\Lambda}}{r}\qty(m+4\pi r^3p)\qty(3m-r+4\pi r^3 p) \right]K\bigg\}\nonumber\\
        &\times\qty{3m+\frac{1}{2}(l+2)(l-1) r+ 4\pi r^3 p}^{-1}\\[3ex]
        V&= \qty{X + \frac{p'}{r}e^{\Phi-\Lambda}W - \frac{1}{2}(p+\varepsilon)e^{\Phi}H_0} \times \qty{\omega^2(p+\varepsilon)e^{-\Phi}}^{-1}\\[3ex]
        H_1' &= \frac{1}{r}\qty[l+1+\frac{2e^{2\Lambda}}{r}m+4\pi r^2(p-\varepsilon)e^{2\Lambda}]H_1 + \frac{e^{2\Lambda}}{r}\qty[H_0+K-16\pi(p+\varepsilon)V]\\[3ex]
        K' &= \frac{H_0}{r}+ \frac{1}{2r}l(l+1)H_1- \qty[\frac{1}{r}(l+1)-\Phi']K - \frac{8\pi}{r}(p+\varepsilon)e^{\Lambda}W\\[3ex]
        W' &= re^{\Lambda}\qty[\frac{e^{-\Phi}}{(p+\varepsilon)}\dv {\varepsilon}{p}X- \frac{1}{r^2}l(l+1)V+\frac{1}{2}H_0+K] -\frac{1}{r}(l+1)W \\[3ex]
        X' &= -\frac{1}{r}lX+ (p+\varepsilon)e^{\Phi}\bigg\{\frac{1}{2}\qty[\frac{1}{r}-\Phi']H_0 + \frac{1}{2}\qty[r\omega^2 e^{-2\Phi}+ \frac{1}{2r}l(l+1)]H_1\nonumber\\
        &\quad - \frac{1}{r}\qty[4\pi(p+\varepsilon)e^{\Lambda}+ \omega^2e^{\Lambda-2\Phi}-r^2\qty(\frac{e^{-\Lambda}}{r^2}\Phi')']W + \frac{1}{2}\qty[3\Phi'-\frac{1}{r}]K- \frac{1}{r^2}l(l+1)\Phi'V \bigg\}
    \end{align} \label{fluid_eq_GR}   
\end{subequations}
The system of differential and algebraic equations, $\ref{fluid_eq_GR}$ completely describes the perturbations inside the star. The differential equations to be solved can be stored in an array $Y= \{H_{1}, K, W, X\}$. This system is clearly singular at $r=0$ and numerically, it will blow up at values of $r$ close to $0$. Thus near the center, $Y(r)$ is approximated as $Y(r)= Y(0)+ \frac{1}{2} Y''(0)r^2+ \mathcal{O}(r^4)$ and the various terms of this approximation are given in \cite{lujunli_ChinPhyB}. At the surface of the star, the pressure perturbations, and thus $X$ must be $0$. To solve equation $\ref{fluid_eq_GR}$, we follow the method outlined in \cite{lindblom1983quadrupole}. We start off with 3 linearly independent solutions at the surface, and 2 linearly independent solutions at the center and integrate them to some point inside the star where they get matched. A linear combination of these solutions, with the coefficients obtained after matching provides the true values of $H_1$ and $K$ at the star's surface. These variables and $H_0$ are the only variables defined outside the star, where the perturbation equations reduce to the Zerilli equation \cite{zhaoUniversalRelationsNeutron2022,lujunli_ChinPhyB}:
\begin{equation}
    \dv[2]{Z}{{r^*}}+ \qty[\omega^2- \mathcal{V}(r)]Z= 0 \label{Zerilli_eq}
\end{equation}
where $\mathcal{V}(r^*)$ is the Zerilli potential, 

\begin{equation}
    \mathcal{V}(r)= \qty(1-2b)\frac{2n^2(n+1)+ 6n^2b+18nb^2+ 18b^3}{r^2(n+3b)^2}
\end{equation}

$r^*$ is the tortoise coordinate, $r^*= r+ 2M\ln\qty(r/2M-1)$, $n= (l-1)(l+2)/2$ and $b=M/r$, with $M$ being the total mass of the star. 

In case of a first-order phase transition inside the HS, we impose additional conditions that ensure the continuity of $H_1$, $K$, $W$, and $X$ across the radius of discontinuity \cite{sotaniDensityDiscontinuityNeutron2001}. 

The perturbed metric outside the star describes a combination of outgoing and incoming gravitational waves, which is the general solution to the Zerilli equation. We are interested in the case of purely outgoing waves, representing the quasi-normal modes (QNM) of the star. At the surface of the star, where $r=R$, the fluid variables can be converted to the Zerilli ones using \cite{zhaoUniversalRelationsNeutron2022}:

\begin{subequations}
    \begin{align}
        Z(R)&= \frac{R\qty[k(R)K(R)-H_1(R)]}{k(R)g(R)-h(R)}\\[3ex]
        \dv{Z(r^*)}{r^*} &= \frac{-h(R)K(R)+g(R)H_1(R)}{k(R)g(R)-h(R)}
    \end{align}\label{zerilli_far}
\end{subequations}
Here

\begin{subequations}
    \begin{align}
        g(r)&= \frac{n(n+1)+3nb+6b^2}{n+3b}\\
        h(r)&= \frac{n- 3nb- 3b^2}{(1-2b)(n+3b)}\\
        k(r)&= 1/(1-2b) 
    \end{align}
\end{subequations}

After continuing the integration of the Zerilli Eq. $\ref{Zerilli_eq}$ to sufficiently far away from the star ($\sim 50\omega^{-1}$), the solution can be approximated as a linear combination of incoming and outgoing waves as $Z(r^*)= A_-(\omega)Z_-(r^*)+ A_+(\omega)Z_+(r^*)$ where $Z_-$ represents the outgoing wave, $Z_+$ the incoming wave and $A_-$ and $A_+$ their amplitudes. At a large enough radius, 

\begin{subequations}
    \begin{align}
        Z_-&= e^{-i\omega r^*} \qty[\beta_0+\frac{\beta_1}{r}+\frac{\beta_2}{r^2}+ \mathcal{O}(r^3)]\\[3ex]
        \dv{Z_-}{r^*}&= -i\omega e^{-i\omega r^*}\qty[\beta_0+ \frac{\beta_1}{r}+ \frac{\beta_2- i\beta_1(1-2M/r)/\omega}{r^2}]
    \end{align}
\end{subequations}
Here $Z_+$ is the complex conjugate of $Z_-$ (and hence $A_+$ the complex conjugate of $A_-$) and, \cite{zhaoUniversalRelationsNeutron2022}
\begin{subequations}
    \begin{align}
        \beta_1&= \frac{-i(n+1)\beta_0}{\omega}\\
        \beta_2&= \frac{\qty[-n(n+1)+ iM\omega(3/2+ 3/n)]\beta_0}{2\omega^2}
    \end{align}
\end{subequations}
$\beta_0$ can be any complex number that represents an overall phase. By matching the solution of $Z(r_*)$ and $\dv{Z(r^*)}{r^*}$ obtained from $\ref{zerilli_far}$ with the above equation, we can find the amplitude $A_+$ with a simple matrix inversion \cite{zhaoUniversalRelationsNeutron2022}. The frequency of the QNM corresponds to that $\omega$ which gives $A_+=0$. 

To find the QNM frequency and its damping time we first find $A_+$, which in general will be a complex number, for several real values of $\omega$ close to the original guess. We then perform a complex polynomial fitting to approximate a parabola passing through the $A_+$ points corresponding to the $\omega$ values. The root of this parabola which has a positive imaginary part is our QNM. We then take the real part of this $\omega$ and repeat the entire procedure several more times till the desired tolerance is reached. The real part of this final $\omega$ is the frequency of the QNM. The inverse of the imaginary part is the corresponding damping time. 
 
As with the Cowling case, the validity of our code was checked thoroughly by comparing our results with those in \cite{2022PhRvD.106f3005K,sotaniDensityDiscontinuityNeutron2001}. Moreover, we also calculated the oscillation modes and damping times using Thorne \cite{1967ApJ...149..591T} Ferrari's \cite{chandrasekhar1991} Breit-Wigner resonance fitting approach and although the damping times could not be obtained with good accuracy for all cases, the frequency values matched exactly to those obtained by the Lindblom \cite{lindblom1983quadrupole,detweiler1985nonradial} approach.
\begin{figure*}[t]
    \begin{subfigure}{0.55\textwidth}
        \centering
        \includegraphics[width=\linewidth]{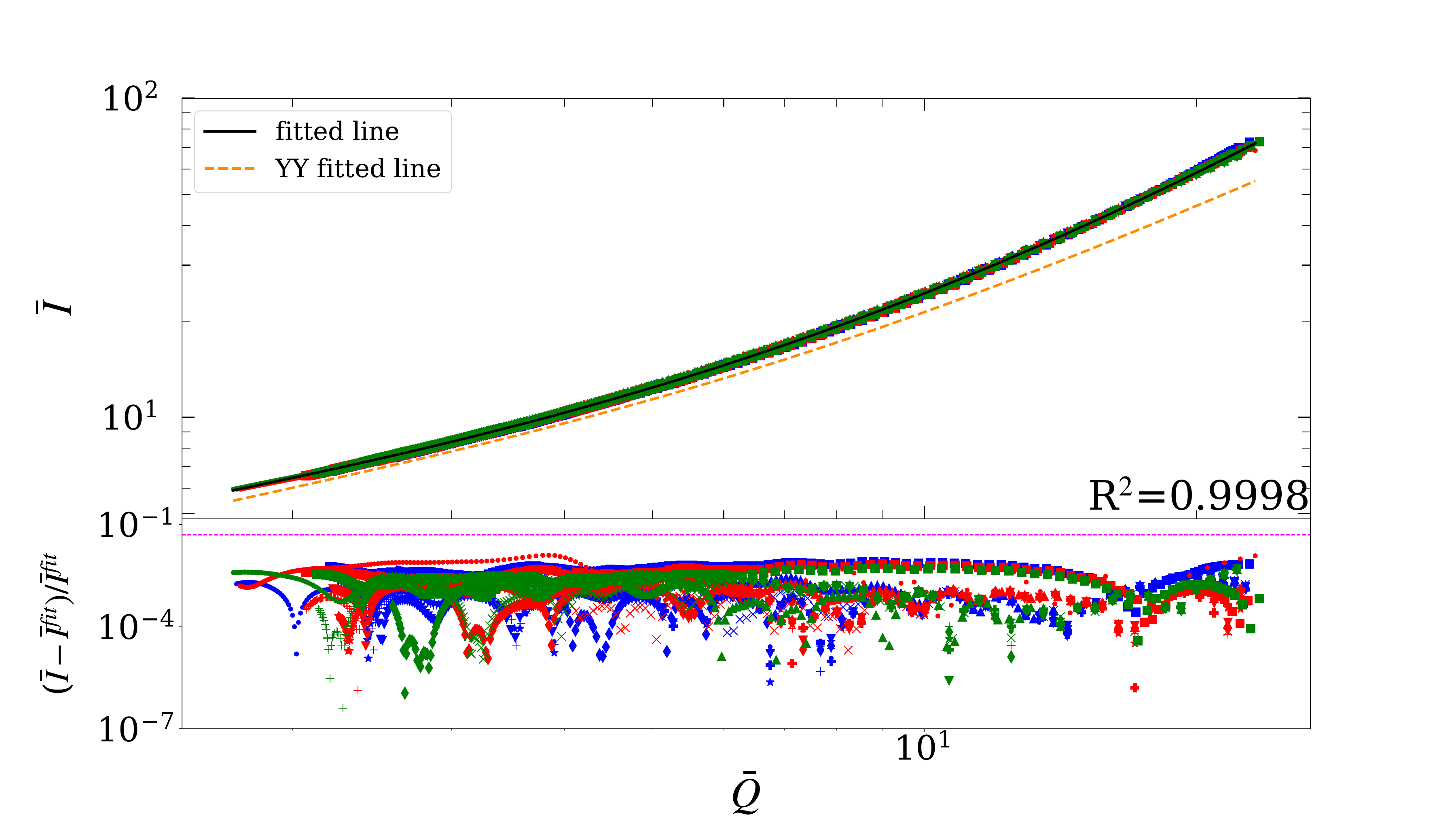}
    \end{subfigure}
    \hspace{-0.85cm} % Added negative space
    \begin{subfigure}{0.49\textwidth}
        \centering
        \includegraphics[width=\linewidth]{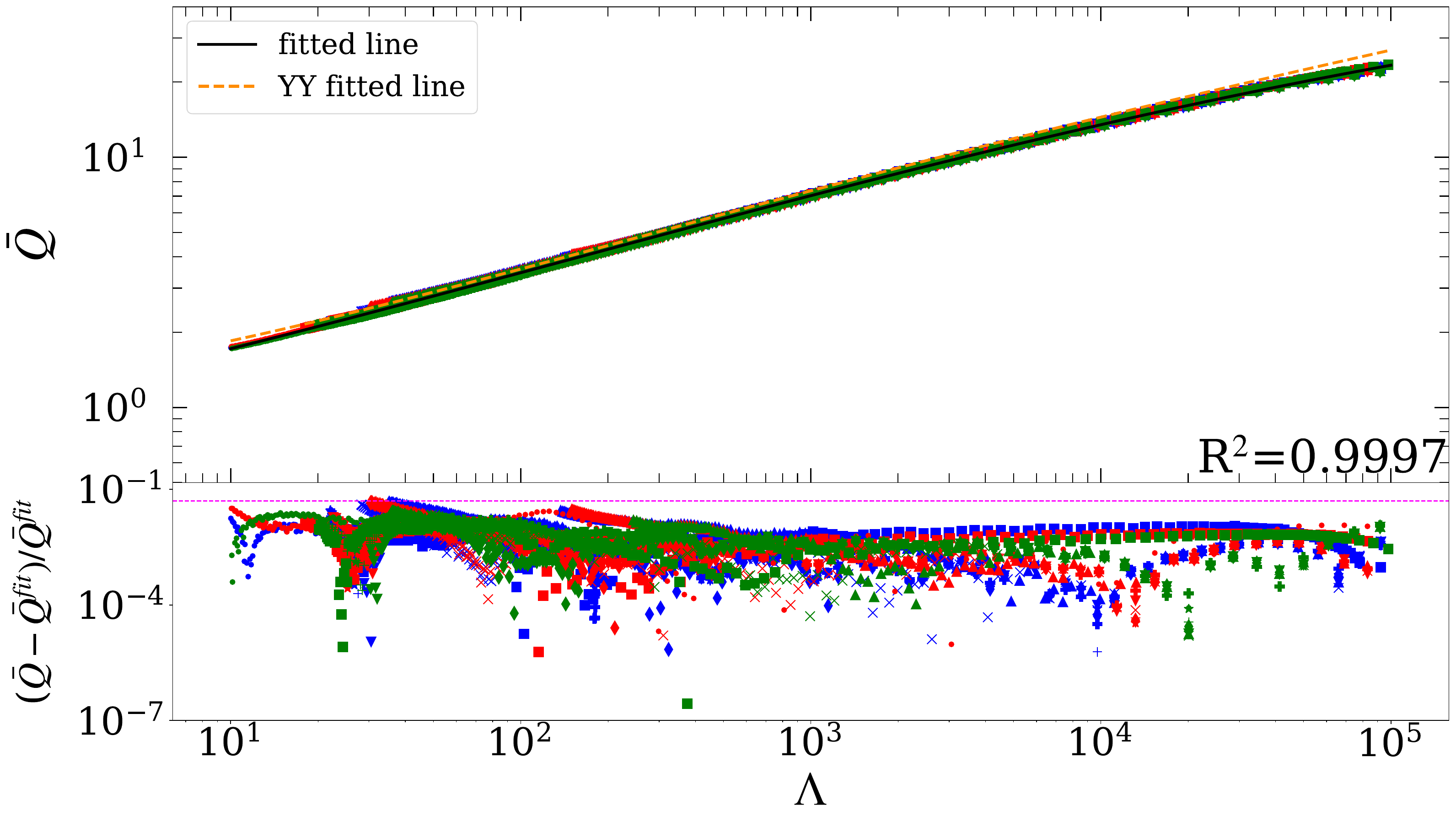}
    \end{subfigure}
    \vskip\baselineskip
    \begin{subfigure}{0.55\textwidth}
        \centering
        \includegraphics[width=\linewidth]{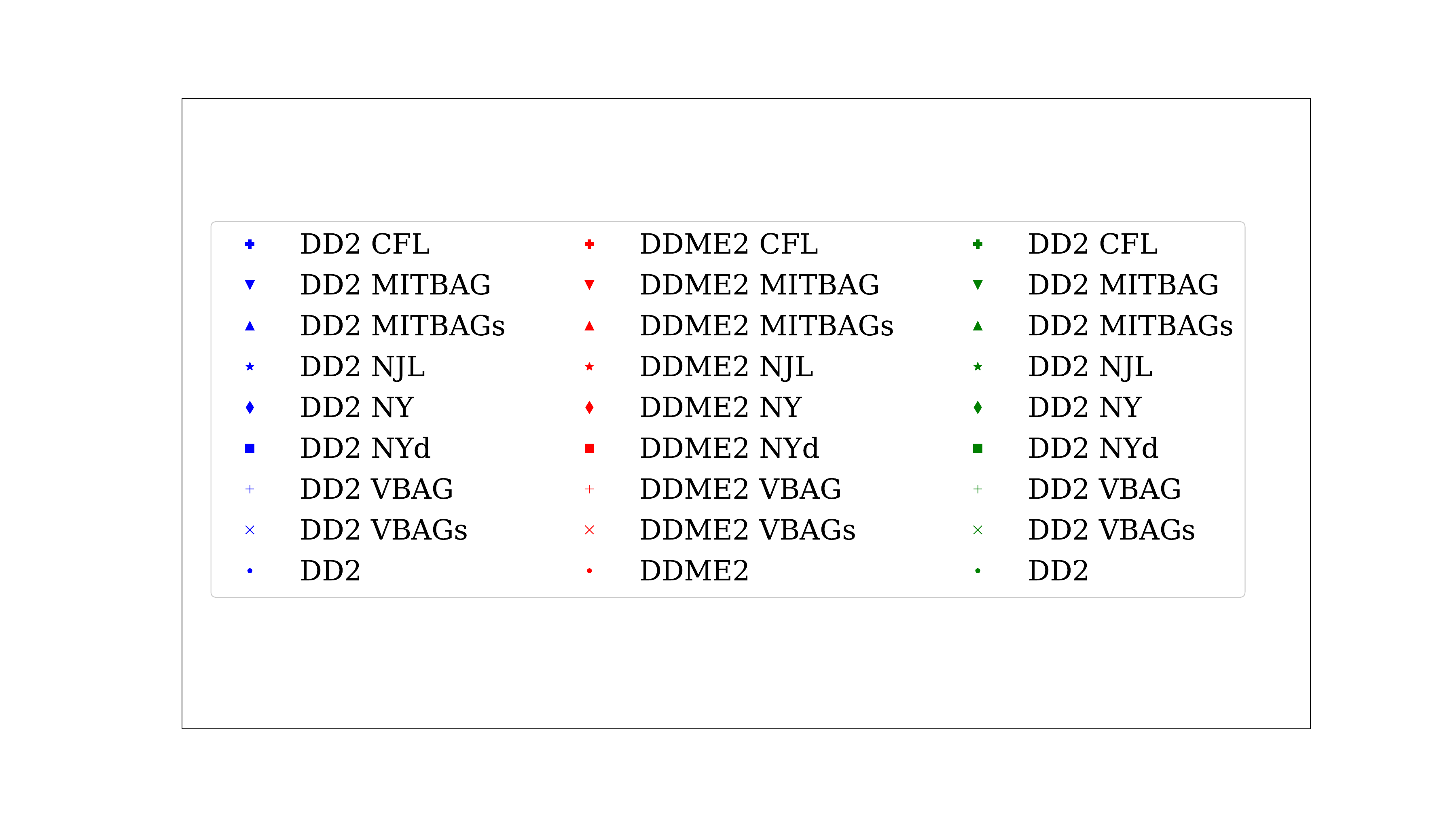}
    \end{subfigure}
    \hspace{-0.85cm}
    \begin{subfigure}{0.49\textwidth}
        \centering
        \includegraphics[width=\linewidth]{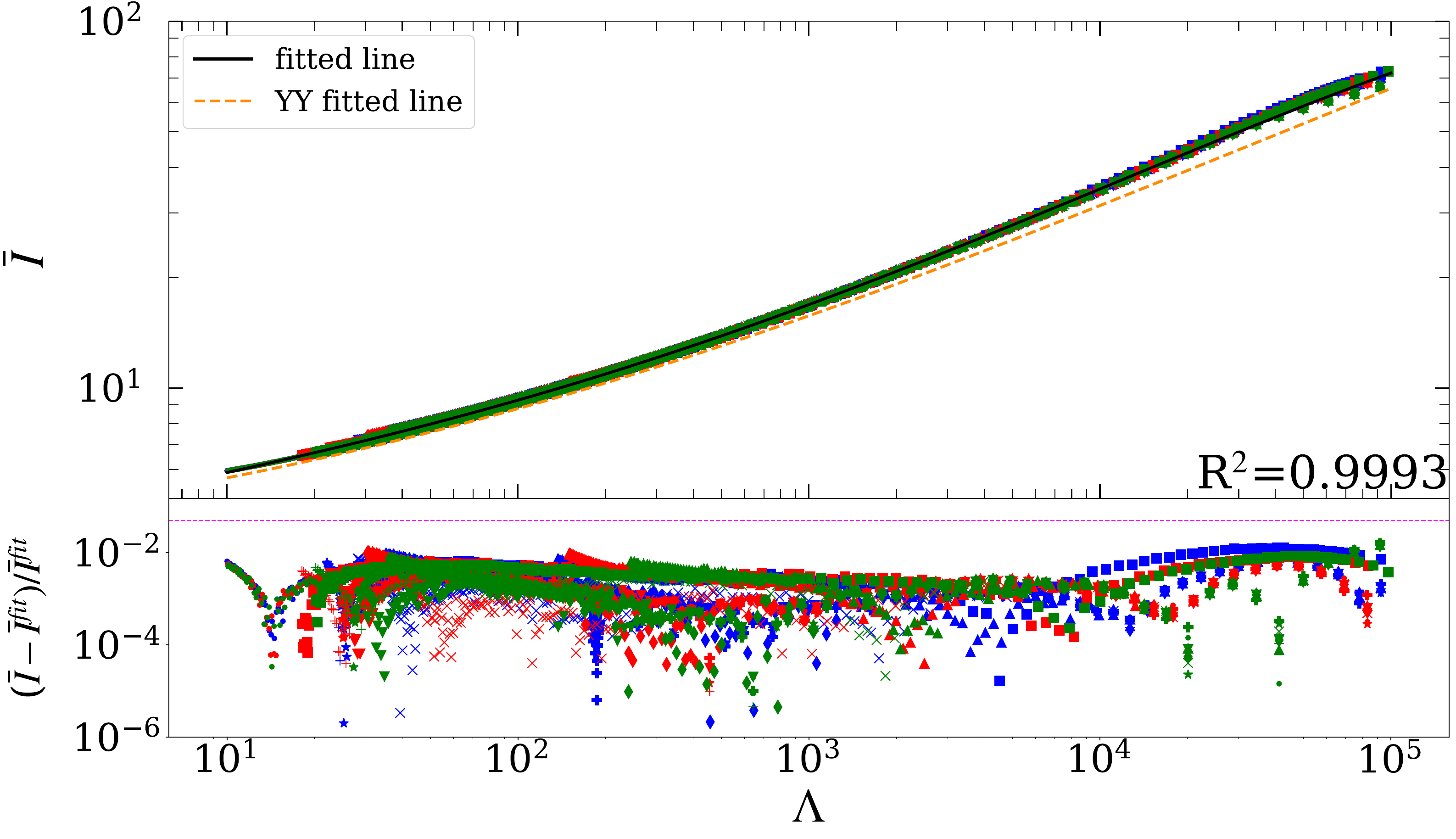}
    \end{subfigure}
    \caption{$I-\Lambda-Q$ relations along with the analytic fit and the fractional error %in the moment of inertia of HSs 
    from the fitting function. Upper left panel: for $I-Q$ relation, upper right panel: for $Q-\Lambda$ relation, and lower right panel: for $I-\Lambda$ relation.} The solid line represents the fitted curve for all data points. The sub-figures display error bars, with the dashed line indicating a $5\%$ error. The nomenclature for the shapes is depicted in the lower left panel.
    \label{fig:universal_plot}
\end{figure*}
\section{Universal relations}\label{sec:universal}
With the CS models discussed above, we study the universal relations for the CSs. First, we study the I-Love-Q relations. For that, as usual, we compute the dimensionless moment of inertia $\Bar{I} = I/M^3$, dimensionless quadrupole moment $\Bar{Q}=-Q/M^3\chi_L^2$, and dimensionless tidal love number or tidal deformability $\Lambda=\Bar{\lambda} = \lambda/M^5$. Here, $I$ is the moment of inertia, $Q$ the spin-induced quadrupole moment \cite{1967ApJ...150.1005H, 2012PhRvL.108w1104P} and $\chi_L=J/M^2$ (J being total angular moment) the dimensionless angular momentum. We compute these quantities for CSs with pure nucleonic matter, hyperonic matter and for HS with normal and superconducting SQM at the core with the help of %These quantities are determined by modeling the HS using 
RNS code \cite{1995ApJ...444..306S, refId0}, which is built on solutions of Einstein’s field equations for axially symmetric and stationary space-time in spherical coordinates. %In our calculation, we assume the star’s 
In this work, we choose the minimum possible frequency of rotation frequency for all the EOSs to find the solutions $\Omega \sim 480$ Hz which is %to be 
much below the mass shedding frequency.

We show the universality in $\bar{I}-\bar{Q},~\bar{Q}-\Lambda$ and $\bar{I}-\Lambda$ plot along with the error due to polynomial fitting various possible composition and parametrization of matter EOS as discussed above in different panels of Fig. \ref{fig:universal_plot}.  The dashed line represents the fitted line by Yagi and Yunes (YY) \cite{2013Sci...341..365Y}. %The plot of the dimensionless moment of inertia ($\Bar{I}$) and the dimensionless quadrupole moment ($\bar{Q}$) is shown in Fig. \ref{fig:universal_IQ_plot}  for $\Bar{I}$ . 
The effectiveness of curve fitting can be evaluated using the coefficient of determination 
\begin{equation}
    R^2 = 1 - \frac{\sum(y_{obs} - y_{expt})^2}{\sum(y_{obs}-y_{mean})^2}
\label{eqn:r_square}    
\end{equation} 

For $I-Q$ relation, shown in the upper left panel of Fig. $\ref{fig:universal_plot}$, we get the value of $R^2=0.9998$ very close to $1$, indicating a good fitting. The maximum deviation, being less than $5\%$, indicates the universality of the $I-Q$ relationship. All the $\bar{I}$ values depend on $\bar{Q}$ rather than properties of matter as expected in a universal relation. The YY fitted line shows some deviation from our fitted line at low density due to a difference in rotation speed \cite{2023PhRvC.108b5810L}. The coefficients for the polynomial fit are given in Table $\ref{tab:fitting_table}$. The variation of $\bar{Q}$ with ${\Lambda}$ is shown in the upper right panel of Fig. $\ref{fig:universal_plot}$. We achieve a good fit with $R^2 = 0.9997$, indicating a high degree correlation. Despite a maximum deviation of approximately $4.65\%$ with the matter composition of nucleonic matter with hyperon in DDME2 parametrization, though it remains below the $5\%$, these findings suggest a good universal relationship between these parameters. Here, the YY fitted line is very close to our fitted line. Similarly, we plot the relation between $\Bar{I}$ and ${\Lambda}$ in the lower right panel of Fig. $\ref{fig:universal_plot}$. We find $R^2 = 0.9993$ and the maximum deviation is $1.56\%$ at low density, which means these parameters are strongly correlated. Our fitted line is almost parallel and close to the YY fitted line.

The universal $I-\Lambda-Q$ relations have been observed in previous literature \cite{2018MNRAS.475.4347R,2023PhRvC.108b5810L} for CSs with nuclear matter, nuclear hyperonic matter, nuclear hyperon-$\Delta$ matter and SQM with NJL model. We observe here that the same universality is applicable for HSs with SQM in the MIT bag model with ad-hoc and vector interaction as well as the CFL phase at the core surrounded by the nucleonic matter.
\begin{figure*}[t]
    \begin{subfigure}{0.54\textwidth}
        \centering
        \includegraphics[width=\linewidth]{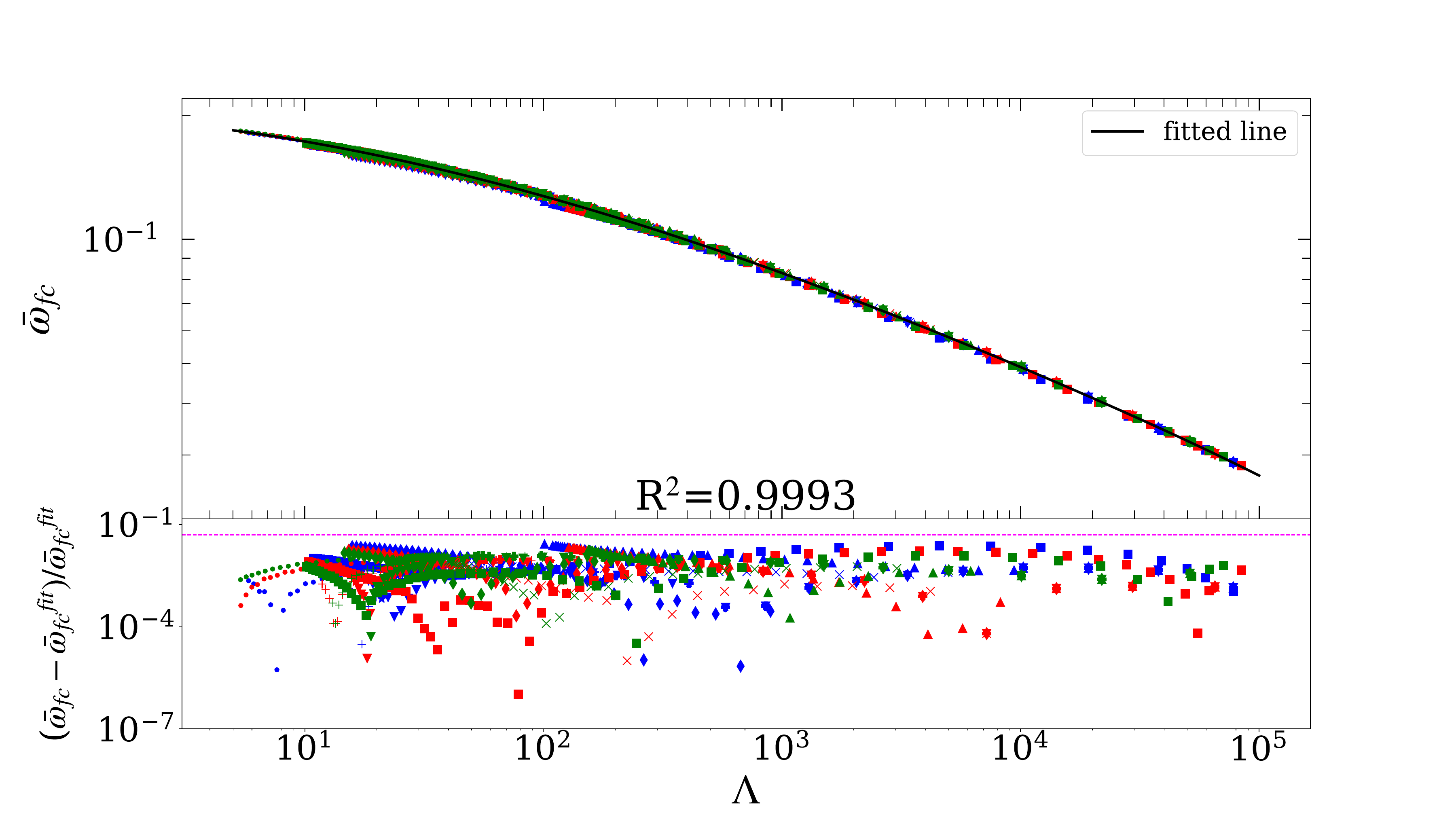}
    \end{subfigure}
    \hspace{-0.8cm} % Added negative space
    \begin{subfigure}{0.49\textwidth}
        \centering
        \includegraphics[width=\linewidth]{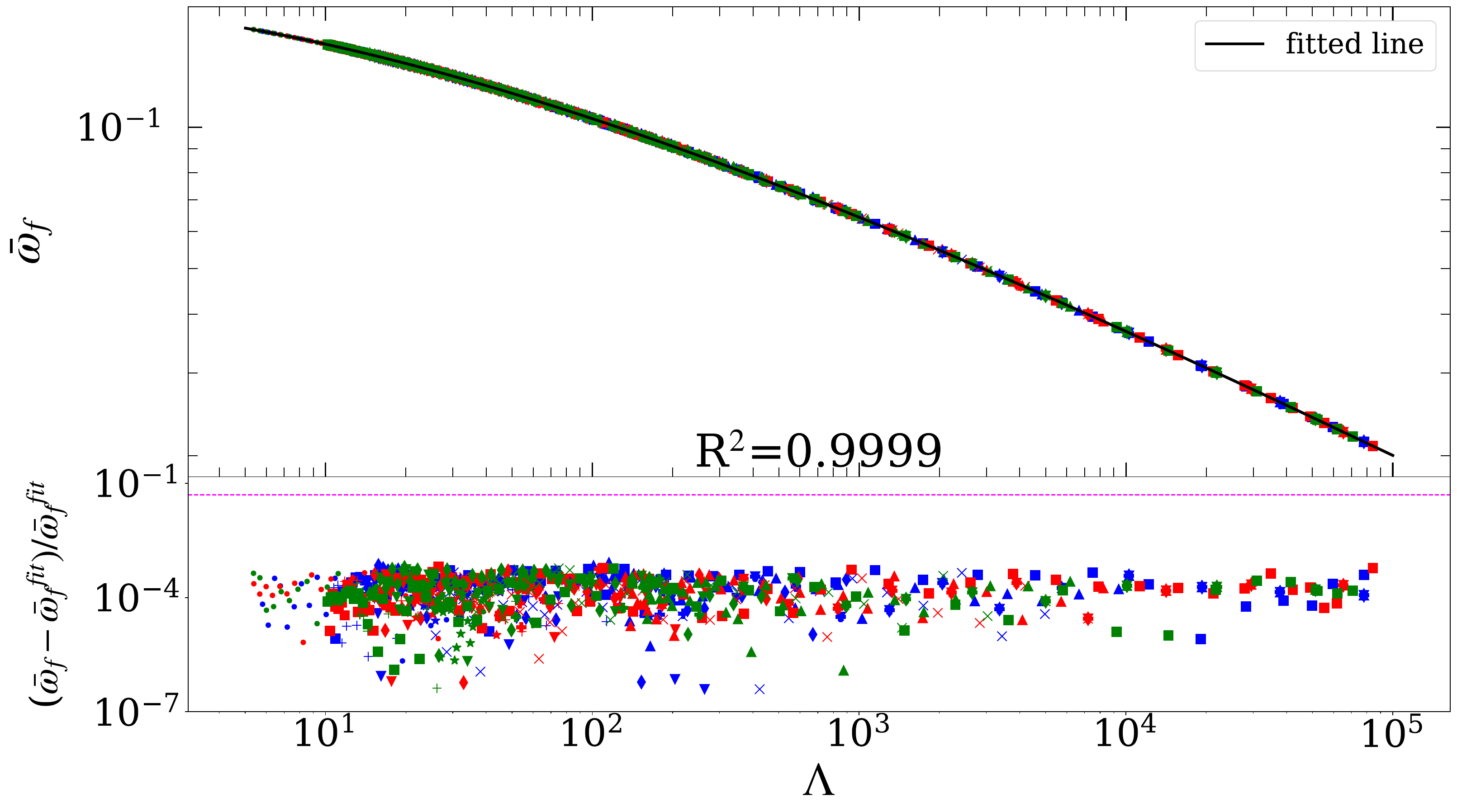}
    \end{subfigure}
    \vskip\baselineskip
    \begin{subfigure}{0.49\textwidth}
        \centering
        \includegraphics[width=\linewidth]{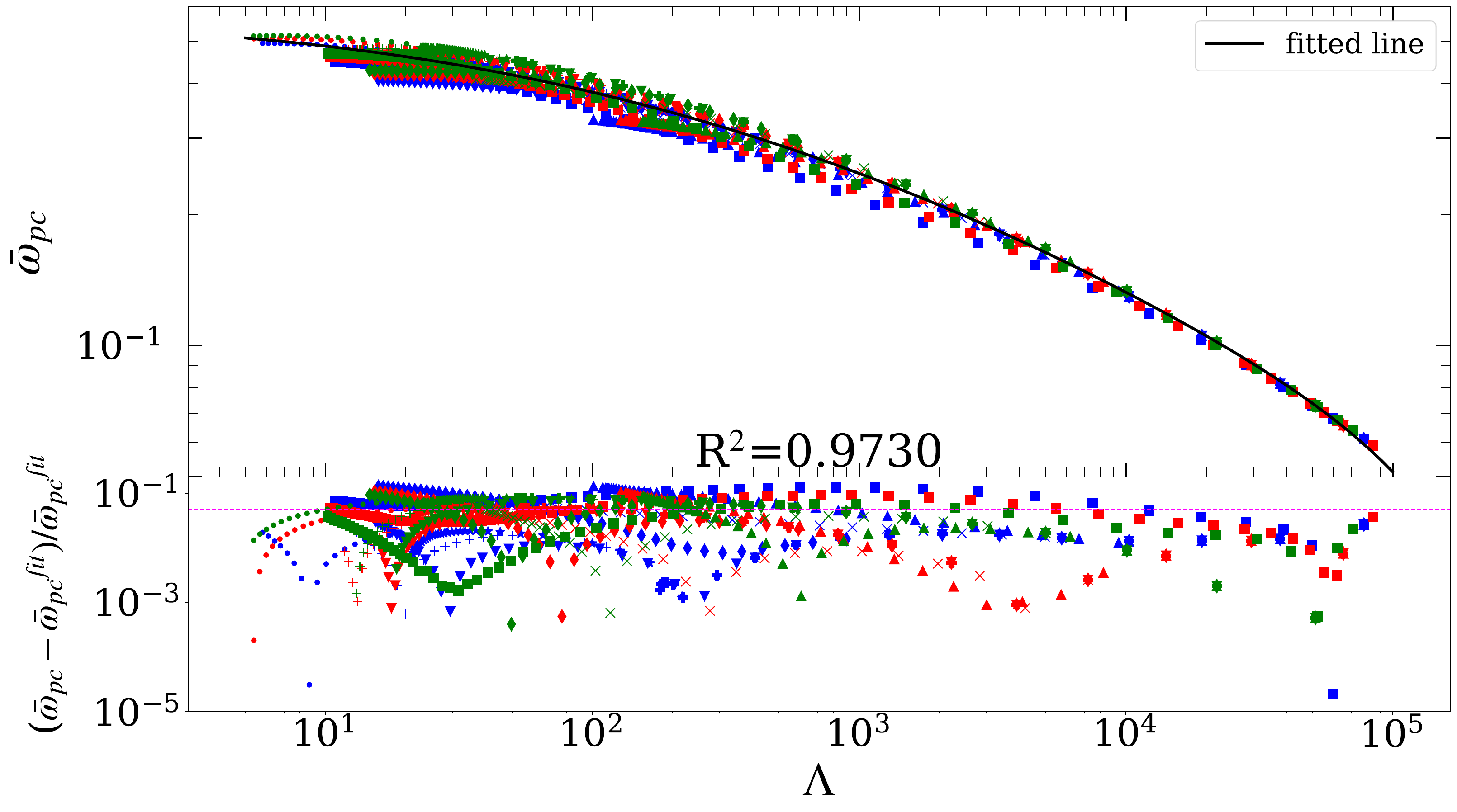}
    \end{subfigure}
    \hspace{0.1cm} % Added negative space
    \begin{subfigure}{0.49\textwidth}
        \centering
        \includegraphics[width=\linewidth]{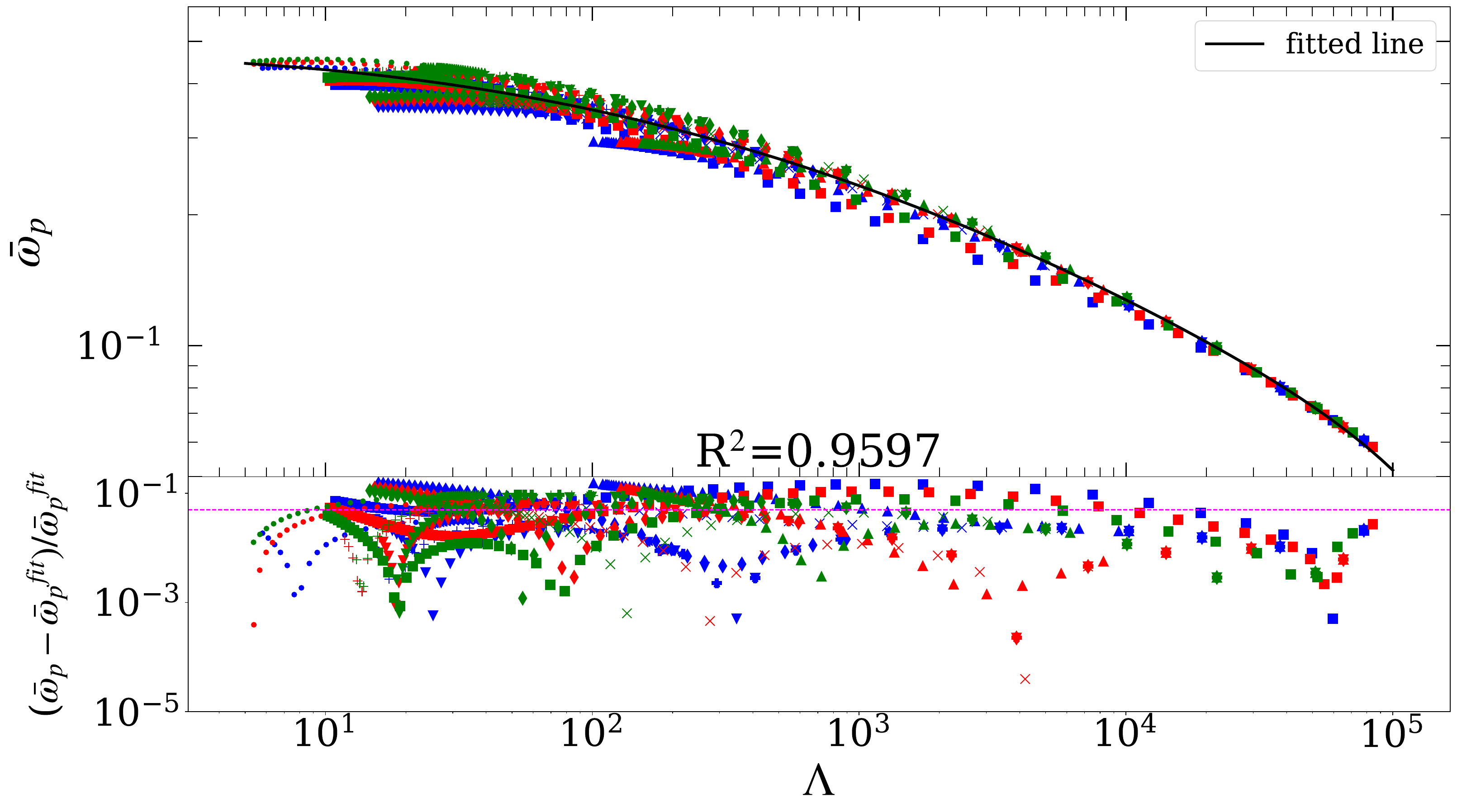}
    \end{subfigure}
    \caption{The relation of dimensionless frequency with tidal deformability. Upper left panel: for f-mode frequency calculated by Cowling approximation and upper right panel: with full GR framework. Similarly, lower panels are for p-mode. The solid line represents the fitted curve for all data points. The sub-figures display error bars, with the dashed line indicating a $5\%$ error. The nomenclature for the shapes remains the same as the Fig. $\ref{fig:universal_plot}$.}
    \label{fig:freq_vs_lam}
\end{figure*}

\begin{figure*}[t!]
    \centering
    
    \begin{subfigure}[h]{0.49\textwidth}
        \centering
        \includegraphics[width=\linewidth]{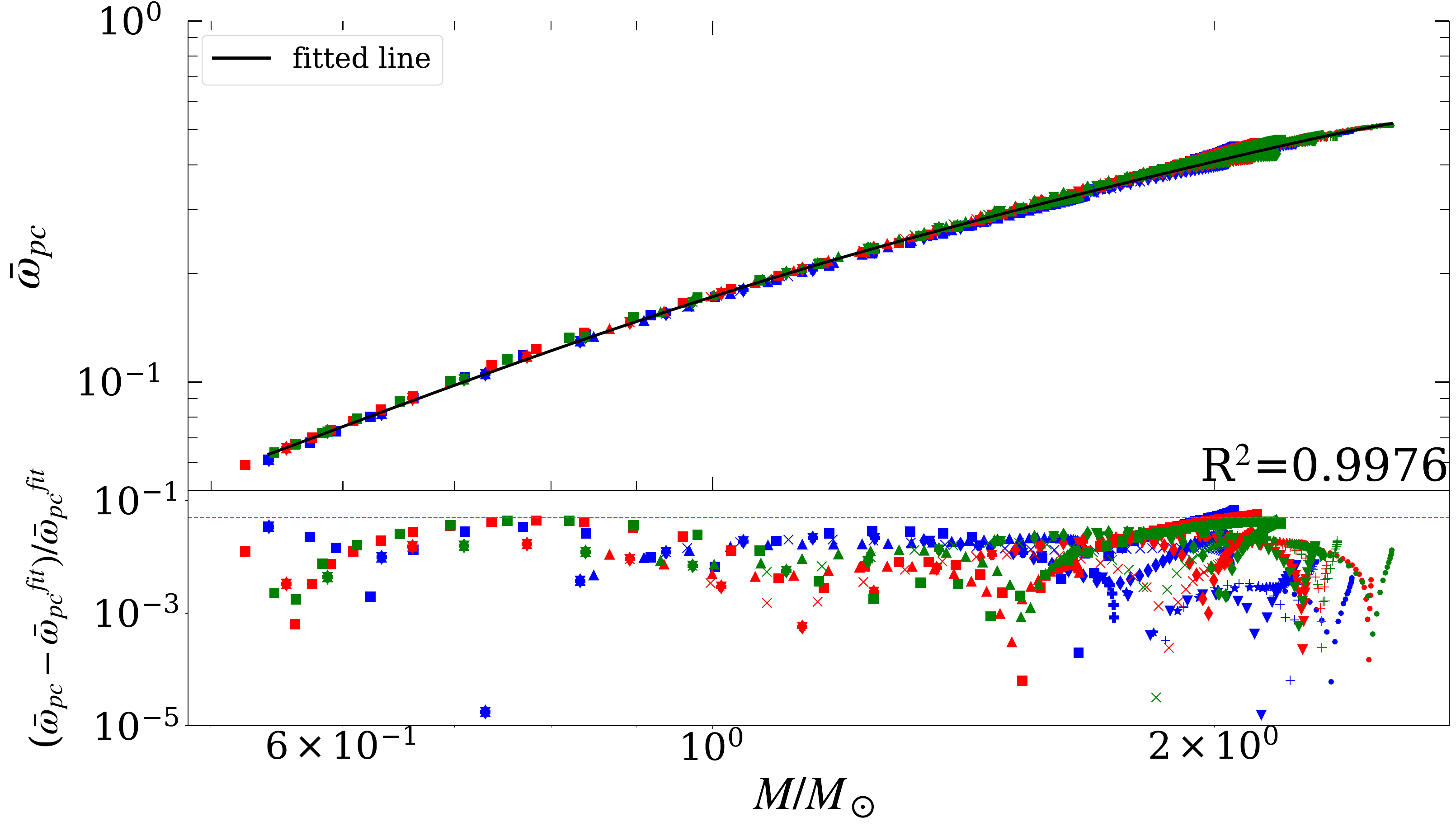}
    \end{subfigure}
    \hfill
    \begin{subfigure}[h]{0.49\textwidth}
        \centering
        \includegraphics[width=\linewidth]{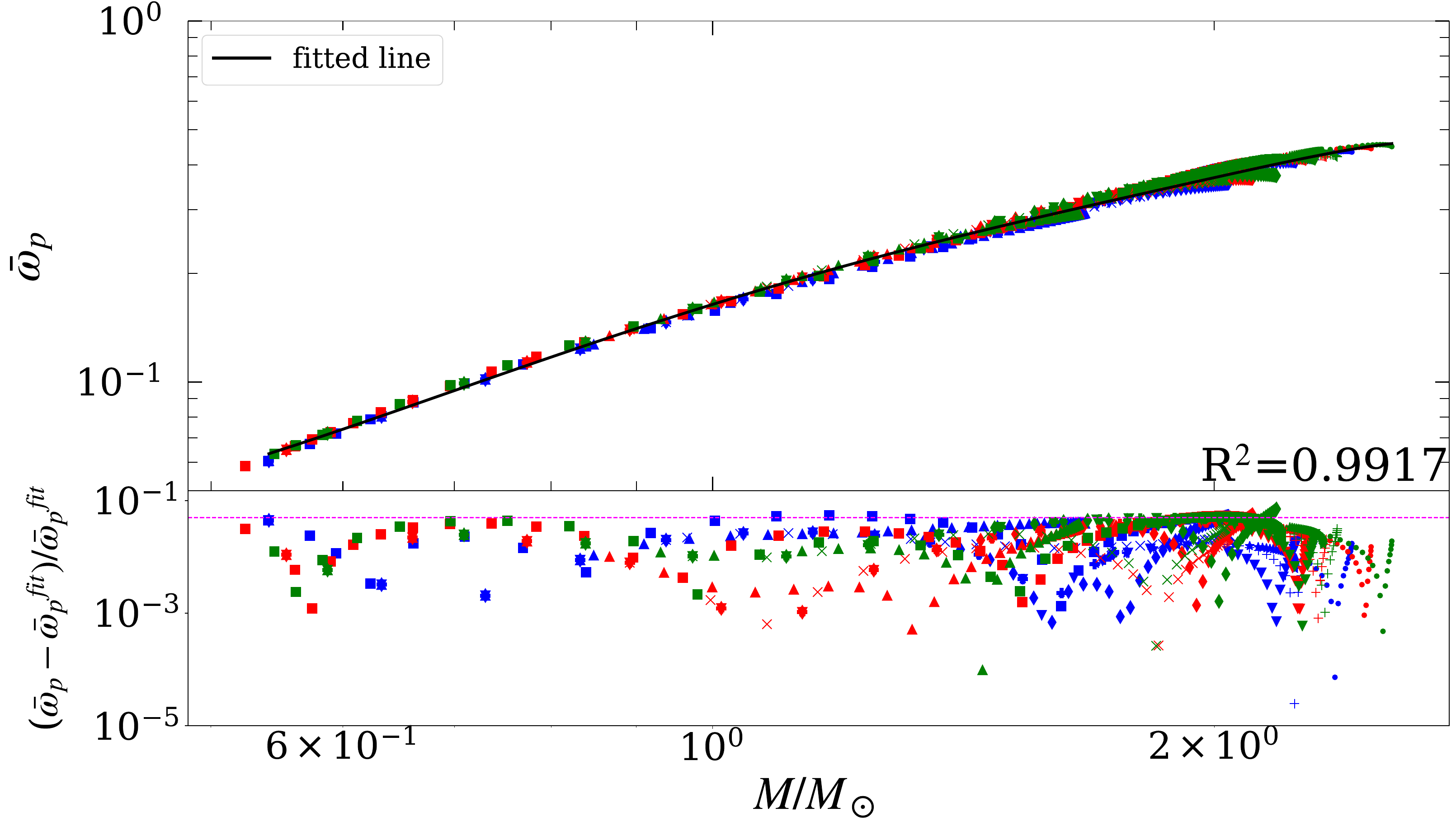}
    \end{subfigure}
    
    \caption{The relation of dimensionless p-mode frequency with star's mass. Left panel: for frequency calculated by Cowling approximation and Right panel: with full GR framework. The solid line represents the fitted curve for all data points. The sub-figures display error bars, with the dashed line indicating a $5\%$ error. The nomenclature for the shapes remains the same as the Fig. $\ref{fig:universal_plot}$.}
    \label{fig:pvsM_plot}
\end{figure*}

\begin{figure*}[h]
    \centering
    
    \begin{subfigure}{0.49\textwidth}
        \centering
        \includegraphics[width=\linewidth]{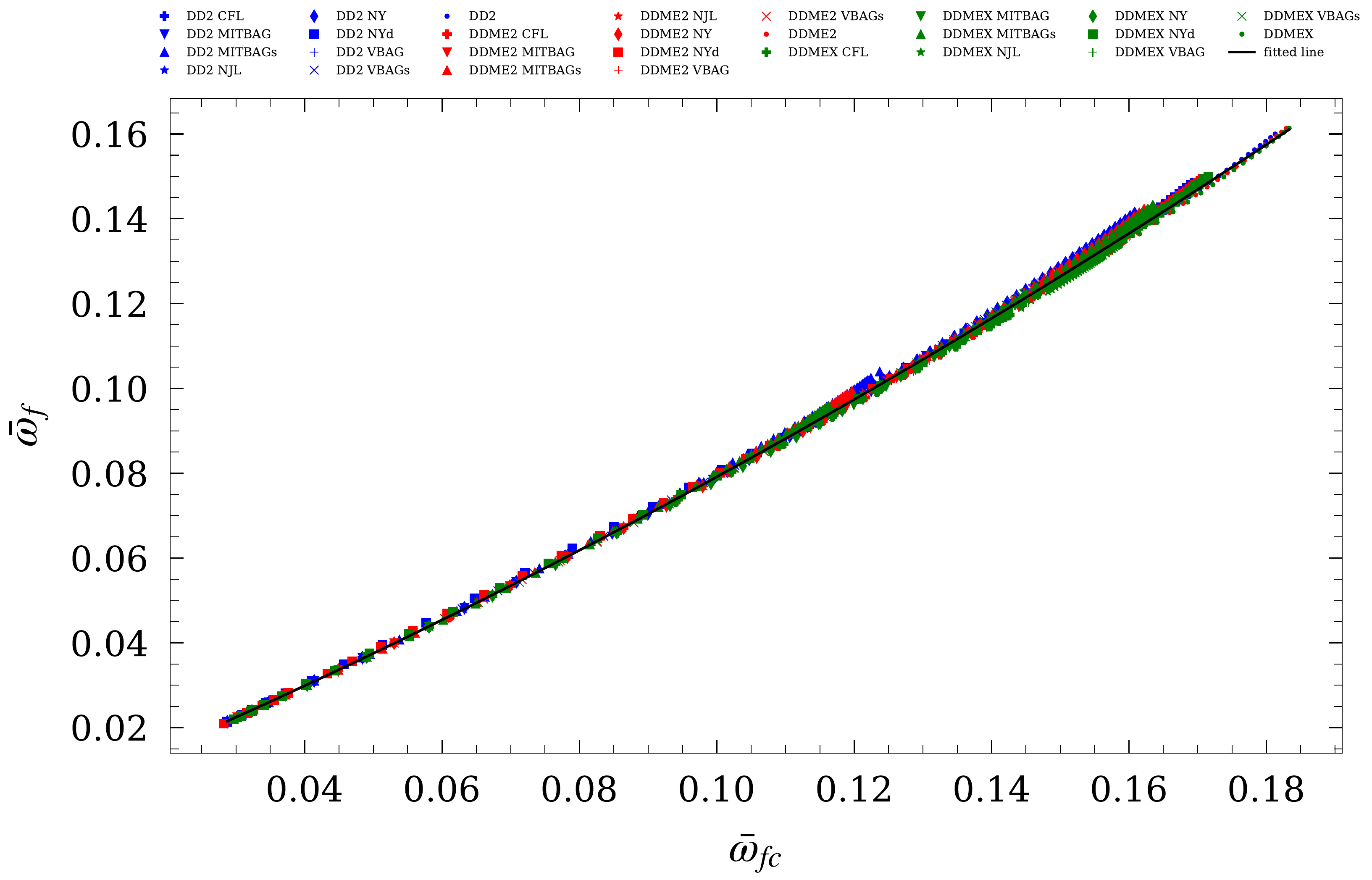}
    \end{subfigure}
    \hfill   
    \begin{subfigure}{0.49\textwidth}
        \centering
        \includegraphics[width=\linewidth]{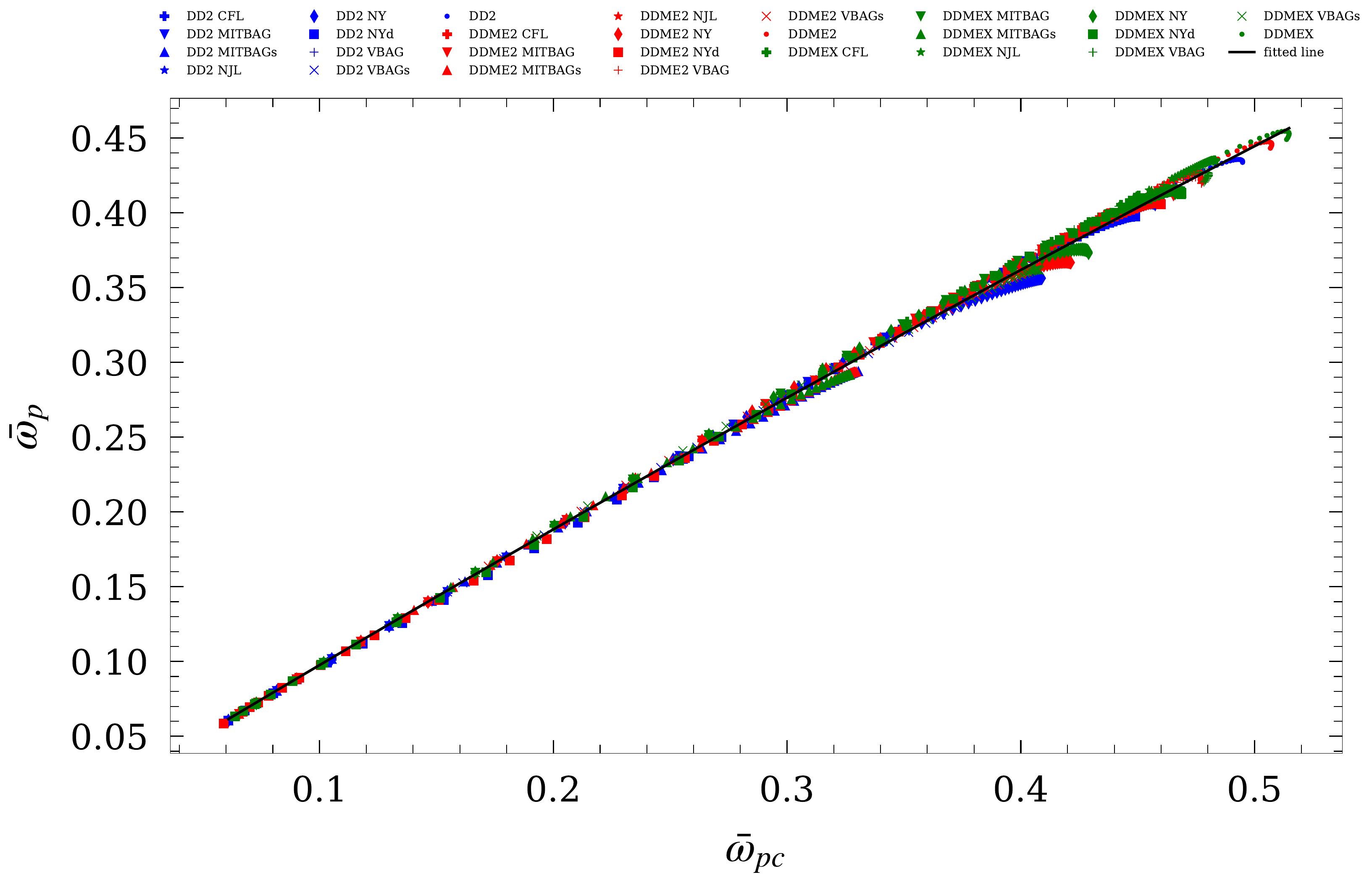}
    \end{subfigure}
    
    \caption{Variation of frequency calculated in full GR framework with frequency calculated by Cowling approximation. Left panel: for f-mode and right panel: for p-mode. The solid line depicts the fitted curve for all data points. The nomenclature for the shapes remains the same as the Fig. $\ref{fig:universal_plot}$.}
    \label{fig:cowl_vs_GR_plot}
\end{figure*}

Next, we focus on correlations in QNMs of non-radial oscillation. We show the relationship of the dimensionless angular frequency $\bar{\omega} = 2{\pi}GM\nu/c^3$ of non-radial oscillations in f- and p-modes with $\Lambda$. Here, $M$ is the star's mass, and $\nu$ is the frequency of the respective non-radial modes in Hz. 

We plot the variation of f-mode frequencies $\bar\omega_{fc}$ using Cowling approximation and $\bar\omega_f$ with $\Lambda$ using full GR in the upper left and right panels respectively along with the analytic fit and fractional error in the angular frequencies $\bar\omega_{fc}$ and $\bar{\omega}_f$ in Fig. \ref{fig:freq_vs_lam}. The value of $R^2$ with Cowling approximation is $0.9993$ and the maximum deviation is $2.65\%$ with DD2 MITBAGs near the maximum mass. If we calculate f-mode frequency in a fully general relativistic framework, this becomes a robust universal relation. Now, $R^2$ increases to the value $0.9999$, which is very close to 1, and the maximum deviation reduces to $0.07\%$. The strength of this correlation has been previously investigated in the references with only pure nucleonic matter and SQM \cite{zhaoUniversalRelationsNeutron2022,PhysRevD.104.123002}.    

Next, we plot the variation of p-mode frequencies $\bar{\omega}_{pc}$ using Cowling approximation and $\bar\omega_p$ with $\Lambda$ using full GR in the lower left and right panels respectively along with the analytic fit and fractional error in the angular frequencies $\bar{\omega}_{pc}$ and $\bar{\omega}_p$ in the lower panels of the same figure. Here, the maximum deviation is about $13.2\%$ from the fitted line with $R^2=0.9730$. The low value of $R^2$ and the high value of maximum deviation suggest that it's a poor fit. Further, we plot p-mode frequency with full GR calculations in the lower panel. Now, the value of $R^2$ further drops to $0.9597$, and the maximum deviation increases to approximately $15\%$. This means the inclusion of perturbation due to gravitational potential or full GR calculation makes it a poorer fit. However, the variation of p-mode frequencies $\bar\omega_{pc}$ using Cowling approximation and $\bar\omega$ using full GR with the stellar mass shows better correlations. This is evident from the left and right panels of Fig. $\ref{fig:pvsM_plot}$. In this case, we obtain the value of $R^2=0.9976$ with Cowling approximation and $R^2=0.9917$ with full GR calculations. These values of $R^2$ indicate a better fit as compared to variation with $\Lambda$. The maximum deviation also drops to  $6.68\%$ with Cowling approximation and $7.07\%$ with full GR. It seems that the p-mode frequency of oscillation is sensitive to matter composition rather than the structure of the star.

If we examine the correlation between the frequencies of non-radial oscillation from Cowling approximation and full general relativistic calculations, we observe, that the effect of metric perturbation increases with stellar mass as expected. Moreover, for f-mode frequency, the effect of metric perturbation is almost independent of matter composition while for p-mode frequency, the metric perturbation effect does depend on the matter composition. This is evident from the left and right panels of Fig. $\ref{fig:cowl_vs_GR_plot}$. This is because the p-modes are more sensitive to the matter distribution inside the star \cite{1998MNRAS.299.1059A}, possibly because of a radial node. The position of the node would be affected by the distribution of matter, which is in turn affected by the back-reaction of the perturbing spacetime on the star. It is worth mentioning that the p-mode frequency is sensitive to pressure at low density as the p-mode frequency correlates with pressure and other matter parameters \cite{2022PhRvD.106f3005K}. We plot the relationship between frequency by Cowling approximation and by applying full GR calculations for both f- and p-modes in Fig. $\ref{fig:cowl_vs_GR_plot}$. This explains the poorer correlation for the p-mode oscillation frequency in the case of full general relativistic calculation. 
\begin{table*}[t!]
\centering
\caption{Fitting parameters for different relations.}
\resizebox{\textwidth}{!}{\begin{tabular}{|l|l|l|l|l|l|l|l|l|}
\hline
y   & x     & $a_5$                  & $a_4$                  & $a_3$                  & $a_2$                  & $a_1$                  & $a_0$                     \\ \hline
$\ln{\bar{I}}$ & $\ln{\Lambda}$ & 0                   & $-1.534\cross10^{-4}$ & $3.261\cross10^{-3}$  & $-1.1\cross10^{-2}$   & 0.1804  & 1.38                \\ \hline
$\ln{\bar{I}}$ &  $\ln{\bar{Q}}$  & 0                   & $-7.928\cross10^{-3}$ & $7.359\cross10^{-2}$  & $-0.541\cross10^{-1}$ & 0.5836              & 1.46               \\ \hline
$\ln{\bar{Q}}$ & $\ln{\Lambda}$ & 0                   & $2.58\cross10^{-5}$   & $-1.503\cross10^{-3}$ & $1.831\cross10^{-2}$  & 0.2284  & -0.0647  \\ \hline
${\bar{\omega}}_{fc}$ & $\ln{\Lambda}$ & $2.754\cross10^{-7}$  & $-2.246\cross10^{-5}$ & $5.547\cross10^{-4}$  & $-4.899\cross10^{-3}$ & -0.002612 & 0.198               \\ \hline
${\bar{\omega}}_{f}$  & $\ln{\Lambda}$ & $9.478\cross10^{-7}$  & $-4.442\cross10^{-5}$ & $7.826\cross10^{-4}$  & $-5.397\cross10^{-3}$ & -0.004411 & 0.184               \\ \hline
${\bar{\omega}}_{pc}$ & $\ln{\Lambda}$ & $-6.575\cross10^{-6}$ & $1.875\cross10^{-4}$  & $-1.392\cross10^{-3}$ & $-8.058\cross10^{-4}$ & -0.01681 & 0.542              \\ \hline
${\bar{\omega}}_{p}$  & $\ln{\Lambda}$ & $-6.747\cross10^{-6}$ & $2.007\cross10^{-4}$  & $-1.718\cross10^{-3}$ & $1.440\cross10^{-3}$  & -0.01272 & 0.467             \\ \hline
${\bar{\omega}}_{pc}$ & M     & -0.0196             & 0.1537              & -0.4675             & 0.6714              & -0.2048             & 0.0390           \\ \hline
${\bar{\omega}}_p$  & M     & -0.0300             & 0.2341              & -0.6991             & 0.9739              & -0.4028             & 0.0881      \\ \hline
${\bar{\omega}}_f$  & ${\bar{\omega}}_{fc}$    & 0             & 0              & 0             & 1.128              & 0.6633             & 0.00153      \\ \hline
${\bar{\omega}}_p$  & ${\bar{\omega}}_{pc}$    & 0             & 0              & 0             & -0.1322              & 0.9467             & 0.0043      \\ \hline
\end{tabular}}
\label{tab:fitting_table}
\end{table*}
For relations we discussed above, we use the following polynomial for curve fitting
\begin{equation}
    y =  a_5x^5 + a_4x^4 + + a_3x^3 + a_2x^2 + a_1x + a_0 
\end{equation}
The coefficients for the curve fitting are given in Table \ref{tab:fitting_table}.
The components of 4th-order are found to be consistent with the ones reported in ref. \cite{2017PhR...681....1Y}. For a bit higher accuracy than the 4th-order, we approach the 5th-order polynomial component fit in cases of non-radial oscillations. The contribution from the 5th-order component can be seen in Table \ref{tab:fitting_table}. The higher degree of polynomials can also be taken for a little bit more accuracy like Zhao et al. \cite{zhaoUniversalRelationsNeutron2022} but the coefficient of the 6th-order becomes very small.
\section{Conclusions and summary}
The CS born after the supernova explosion contains matter at a density a few times $n_0$. At that high density, the matter properties are not well constrained from experimental as well as theoretical points of view. Consequently, the matter at high density is modeled for different compositions and interactions between constituent particles. The macroscopic properties of CS are highly sensitive to highly dense matter models. However, a few relations between some macroscopic quantities are independent of microscopic models of highly dense matter. Yagi and Yunes first pointed this out in 2013 \cite{2013Sci...341..365Y, 2013PhRvD..88b3009Y} and further supported by many studies \cite{2016CQGra..33mLT01Y,2018PhRvD..97h4038P,2018CQGra..35a5005S,2018PhRvD..97f4042G,2017PhRvC..96d5806M,2019JPhG...46c4001W,2013PhRvD..88b3007M, 2013ApJ...777...68B, 2014MNRAS.438L..71H, 2014PhRvL.112l1101P, 2014PhRvD..89l4013Y,2014PhRvL.112t1102C,2021MNRAS.502.3476S,2019PhRvD..99d3004R,2020PhRvD.101l4006J,2021PhRvD.103f3036G,2021ApJ...906...98N,2022MNRAS.515.3539K,2022PhRvD.106l3002Z}. Most of the studies of universal relations have been done with stars composed of pure nucleonic matter - the neutron star. Other possible configurations of compact stars are baryonic stars with heavier baryons in the inner part of the star and the HS. The universal relation with HS configuration has also been studied in refs. \cite{2018PhRvD..97h4038P, 2022MNRAS.515.3539K} with Maxwell construction of phase transition to SQM in normal phase. In this current work, we examine the universal relation for HS with Gibbs construction to SQM in the normal phase with MIT bag model with ad-hoc or vector interactions and in the CFL phase and we find that this kind of HS also maintains the same universality with the CS made up of pure baryonic matter. Then we find the universal relations between the frequencies of non-radial f-mode oscillation with dimensionless tidal deformability for the HS configuration which is compatible with the pure baryonic star configuration. However, the frequency of non-radial p-mode oscillation does not correlate well with dimensionless tidal deformability. The correlation improves with mass but is not up to the mark. We find that p-mode oscillation frequency is sensitive to the composition of matter in both Cowling approximation and full general relativistic calculations.

\section*{Data Availability}
The data used in the manuscript can be obtained at reasonable request from the corresponding author.

\section*{Acknowledgements}
The authors are very thankful to B.K. Agarwal for carefully reading the manuscript and the valuable comments in presenting the results. Authors are also thankful to  N. Jokela, and V. Parmar for their helpful comments on Arxiv version 1. The authors acknowledge the financial support from the Science and Engineering Research Board (SERB), Department of Science and Technology, Government of India through Project No. CRG/2022/000069. K. K. Nath would like to acknowledge the Department of Atomic Energy (DAE), Govt. of India, for sponsoring the fellowship covered under the sub-project no. RIN4001-SPS (Basic research in Physical Sciences). The authors thank the anonymous referee for enhancing the manuscript’s quality with valuable comments.   

\bibliographystyle{spphys}       % APS-like style for physics
\bibliography{references}   % name your BibTeX data base
\end{document}